\documentclass{iopart}
\usepackage{graphicx}  
\usepackage{latexsym}
\usepackage{amssymb}

\jl{6}        
\eqnobysec    

\def\beq{\begin{equation}}
\def\eeq{\end{equation}}

\def\rmd{{\rm d}}
\def\version{\today}

\begin{document}

\begin{flushright}
Current version: \version \\
\end{flushright}

\title[Spin precession in the Schwarzschild spacetime: circular orbits]
{Spin precession in the Schwarzschild spacetime: circular orbits}

\author{
Donato Bini$^* {}^\S{}^\P$,
Fernando de Felice$^\dagger$, 
Andrea Geralico$^\ddag {}^\S$
and Robert T. Jantzen$^\diamond {}^\S$
}
\address{
  ${}^*$\
Istituto per le Applicazioni del Calcolo ``M. Picone'', CNR I-00161 Rome, Italy
}
\address{
  ${}^\S$\
  International Center for Relativistic Astrophysics,
  University of Rome, I-00185 Rome, Italy
}
\address{
${}^\P$
  INFN - Sezione di Firenze, Polo Scientifico, Via Sansone 1, 
  I-50019, Sesto Fiorentino (FI), Italy 
}
\address{
${}^\dagger$\
Dipartimento di Fisica, Universit\`a di Padova, and INFN, Sezione di Padova, Via Marzolo 8,  I-35131 Padova, Italy
}
\address{
  ${}^\ddag$\
  Dipartimento di Fisica, Universit\`a di Lecce, and INFN - Sezione di Lecce,
  Via Arnesano, CP 193, I-73100 Lecce, Italy
}
\address{${}^\diamond$
Department of Mathematical Sciences, Villanova University, Villanova, PA 19085, USA
}

\begin{abstract}
We study the behavior of nonzero rest mass spinning test particles moving along circular orbits in the Schwarzschild spacetime in the 
case in which the components of the spin tensor are allowed to vary along the orbit, generalizing some previous work. 
\end{abstract}

\pacno{04.20.Cv}

\section{Introduction}

The  equations of motion  for a spinning test particle in a given gravitational background were deduced
by Mathisson and Papapetrou \cite{math37,papa51} and read
\begin{eqnarray}
\label{papcoreqs1}
\frac{DP^{\mu}}{\rmd \tau_U}
&=&-\frac12R^{\mu}{}_{\nu\alpha\beta}U^{\nu}S^{\alpha\beta}\equiv F^{\rm (sc)}{}^{\mu}
\ , \\
\label{papcoreqs2}
\frac{DS^{\mu\nu}}{\rmd \tau_U}&=&P^{\mu}U^{\nu}-P^{\nu}U^{\mu}\ ,
\end{eqnarray}
where $P^{\mu}$ is the total four-momentum of the particle,  $S^{\mu\nu}$ is the antisymmetric spin tensor, and 
$U$ is the 4-velocity of the timelike ``center of mass line'' used to make the multipole reduction.
In order to have a closed set of equations, Eqs.~(\ref{papcoreqs1})  and (\ref{papcoreqs2}) must be completed 
by adding  supplementary conditions (SC) whose standard choices in the literature are the
\begin{itemize}
\item[1.]
Corinaldesi-Papapetrou \cite{cori51} conditions (CP): $S^{t\nu}=0$,
where the index $\nu$ corresponds to a coordinate component and $t$ is a timelike slicing coordinate,
\item[2.]
Pirani \cite{pir56} conditions (P): $S^{\mu\nu}U_\nu=0$, 
\item[3.]
Tulczyjew \cite{tulc59} conditions (T): $S^{\mu\nu}P_\nu=0$.
\end{itemize}
Only solutions of the combined equations for which both $U$ and $P$ are timelike vectors are considered, in order to have a meaningful interpretation describing a spinning test particle with nonzero rest mass and physical momentum.

Not much is known about actual solutions of these equations in explicit spacetimes which satisfy the Einstein equations.
In a previous article \cite{bdfg1}, we considered the simplest special case of a spinning test particle moving uniformly along a circular orbit in the static spherically symmetric Schwarzschild spacetime, but because these equations are still complicated, we looked for solutions with constant frame components of the spin tensor in the natural symmetry adapted static frame, i.e., coinciding with a static tensor field along the path. Such a static spin tensor is a very strong restriction on the solutions of these equations of motion, leading
to special solutions in which the spin vector is perpendicular to the plane of the orbit,
and contributes to an adjustment in the acceleration of the orbit. 

Here we consider the slightly less restrictive case where the spin components are not constant, but the motion is still circular.
However, in this case it is clear that if the spin tensor has time-dependent components, its feedback into the acceleration of the test particle path will break the static symmetry of that path unless the spin precession is very closely tied to the natural Frenet-Serret rotational properties of the path itself. Indeed we find that only the Pirani supplementary conditions permit such specialized solutions since they allow the spin tensor to be described completely by a spatial spin vector in the local rest space of the path itself. By locking spin vector precession to the Frenet-Serret rotational velocity of the path, solutions are found with a spin vector Fermi-Walker transported along an accelerated center of mass world line. The remaining choices for the supplementary conditions have no natural relationship to the Frenet-Serret properties of the particle path and do not admit such specialized solutions. 

With the assumption of circular motion, one can solve the equations of motion explicitly up to constants of integration. By a process of elimination, one can express them entirely in terms of the spin components and particle mass as a constant coefficient linear system of first and second order differential equations. By systematic solving and backsubstitution, one gets decoupled linear second order constant coefficient equations for certain spin components, which are easily solved to yield exponential or sinusoidal or quadratic solutions as functions of the proper time, from which the remaining variables may be calculated. Imposing the choice of supplementary conditions then puts constraints on the constants of integration or leads to inconsistencies. The details of the decoupling and solution of the equations of motion are left to the Appendix, leaving the imposition of the supplementary conditions to the main text. 

\section{Circular orbits in the Schwarzschild spacetime}

Consider the case of the Schwarzschild spacetime, with the metric written in standard coordinates
\beq\fl\quad 
\label{metric}
\rmd  s^2 = -\left(1-\frac{2M}r\right)\rmd t^2 + \left(1-\frac{2M}r\right)^{-1} \rmd r^2 
+ r^2 (\rmd \theta^2 +\sin^2 \theta \rmd \phi^2)\ ,
\eeq
and introduce the usual orthonormal frame adapted to the static observers following the time lines
\beq\fl\quad 
\label{frame}
e_{\hat t}=(1-2M/r)^{-1/2}\partial_t, \,
e_{\hat r}=(1-2M/r)^{1/2}\partial_r, \,
e_{\hat \theta}=\frac{1}{r}\partial_\theta, \,
e_{\hat \phi}=\frac{1}{r\sin \theta}\partial_\phi ,
\eeq
with dual frame
\begin{equation}\fl\quad 
\omega^{{\hat t}}=(1-2M/r)^{1/2}\rmd t\,, \, 
\omega^{{\hat r}}=(1-2M/r)^{-1/2}\rmd r\,, \, 
\omega^{{\hat \theta}}=r \rmd \theta\,, \,
\omega^{{\hat \phi}}=r\sin \theta \rmd\phi\,,
\end{equation}
where $\{\partial_t, \partial_r, \partial_\theta, \partial_\phi\}$ and $\{\rmd t, \rmd r, \rmd \theta,\rmd \phi\}$ are the coordinate basis and its dual, respectively.

In order to investigate the simplest special solutions of the combined equations of motion,
we explore the consequences of assuming that the test particle 4-velocity $U$ corresponds to a timelike constant speed circular orbit confined to the equatorial plane $\theta=\pi/2$. Then it must have the form
\beq
\label{orbita}
U=\Gamma [\partial_t +\zeta \partial_\phi ]
=\gamma [e_{\hat t} +\nu e_{\hat \phi}], \qquad 
\gamma=(1-\nu^2)^{-1/2}\ ,
\eeq
where $\zeta$  is
the angular velocity with respect to infinity, $\nu$ is the azimuthal velocity as seen by the static observers, $\gamma$ is the associated gamma factor, and $\Gamma$ is a normalization factor which assures that $U\cdot U=-1$. These are related by
\beq\fl\quad
\zeta=(-g_{tt}/g_{\phi\phi})^{1/2} \nu \ ,\qquad
\Gamma =\left( -g_{tt}-\zeta^2g_{\phi\phi} \right)^{-1/2}
  =(-g_{tt})^{-1/2} \gamma 
\ ,
\eeq
so that $\zeta\Gamma = \gamma\nu/(g_{\phi\phi})^{1/2}$, which reduces to 
$\gamma\nu/r$ in the equatorial plane.

Here $\zeta$ and therefore $\nu$ are assumed to be constant along the world line. 
We limit our analysis to  the equatorial plane $\theta=\pi/2$; as a convention, the physical (orthonormal) 
component along $-\partial_\theta$ which is perpendicular to the equatorial plane will be referred to as ``along the positive $z$-axis" and will be indicated by the index $\hat z$ when convenient: $e_{\hat z}=-e_{\hat \theta}$. Note both $\theta=\pi/2$ and $r=r_0$ are constants along any given circular orbit, and that the azimuthal coordinate along the orbit depends on the coordinate time $t$ or proper time $\tau$ along that orbit according to 
\beq\label{eq:phitau}
  \phi -\phi_0 = \zeta t = \Omega_U \tau_U \ ,\quad
\Omega_U =\gamma\nu/r
\ ,
\eeq
defining the corresponding coordinate and proper time orbital angular velocities $\zeta$ and $\Omega_U$. These determine the rotation of the spherical frame with respect to a nonrotating frame at infinity. 

Among all circular orbits the timelike circular geodesics merit special attention, whether co-rotating $(\zeta_+)$
or counter-rotating $(\zeta_-)$ with respect to increasing values of the azimuthal coordinate $\phi$ (counter-clockwise motion). Their time coordinate angular velocities  
$\zeta_\pm\equiv \pm\zeta_K=\pm (M/r^3)^{1/2}$, which are identical with the Newtonian Keplerian values,  lead to the expressions
\beq\fl\quad 
\label{Ugeos}
U_\pm=\gamma_K [e_{\hat t} \pm \nu_K e_{\hat \phi}]\ , \qquad 
\nu_K=\left[\frac{M}{r-2M}\right]^{1/2}\ , \qquad 
\gamma_K=\left[\frac{r-2M}{r-3M}\right]^{1/2}\ ,
\eeq
where the timelike condition $\nu_K < 1$ is satisfied if $r>3M$. At $r=3M$ these circular geodesics go null.

It is convenient to introduce the Lie relative curvature \cite{idcf1,idcf2} of each orbit 
\beq
k_{\rm (lie)}=-\partial_{\hat r} \ln \sqrt{g_{\phi\phi}}=-\frac1r\left(1-\frac{2M}{r}\right)^{1/2}=-\frac{\zeta_K}{\nu_K}\ ,
\eeq
and a Frenet-Serret intrinsic frame along $U$ \cite{iyer-vish}, defined by  
\beq 
\label{FSframe}
E_{0}=U\ , \quad
E_{1}=e_{\hat r}\ , \quad
E_{2}=\gamma[\nu e_{\hat t} +e_{\hat \phi}]\ , \quad
E_{3}=e_{\hat z}
\eeq
satisfying the following system of evolution equations along the constant radial acceleration orbit
\beq
\label{FSeqs}\fl\quad
\frac{DU}{d\tau_U} \equiv a(U)=\kappa E_{1}\ ,\ 
\frac{DE_{1}}{d\tau_U}=\kappa U+\tau_1 E_{2}\ ,\
\frac{DE_{2}}{d\tau_U}=-\tau_1 E_{1}\ ,\  
\frac{DE_{3}}{d\tau_U}=0\ ,
\eeq
where in this case 
\begin{eqnarray}\label{kappatau1}
\kappa &=& k_{\rm (lie)}\gamma^2[\nu^2-\nu_K^2]
 = -\frac{\gamma^2(\nu^2-\nu_K^2)}{\nu_K} \zeta_K
\ , \nonumber\\
\tau_1 &=& -\frac{1}{2\gamma^2} \frac{d\kappa}{d\nu}
=-k_{\rm (lie)}\frac{\gamma^2}{\gamma_K^2}\nu
= -\frac{\gamma^2\nu}{\gamma_K^2\nu_K} \zeta_K
\ . 
\end{eqnarray}

The projection of the spin tensor into the local rest space of the static observers defines the spin vector by spatial duality
\beq
S^\beta=\frac12 \eta_\alpha{}^{\beta\gamma\delta}(e_{\hat t})^\alpha S_{\gamma\delta}\ ,
\eeq
where $\eta_{\alpha\beta\gamma\delta}=\sqrt{-g} \epsilon_{\alpha\beta\gamma\delta}$ is the unit volume 4-form constructed from the Levi-Civita alternating symbol $\epsilon_{\alpha\beta\gamma\delta}$ ($\epsilon_{\hat t\hat r\hat\theta\hat\phi}=1$),
leading to the correspondence
\beq
\label{spinvecehatt}
(S^{\hat r},S^{\hat\theta}=-S^{\hat z},S^{\hat\phi})
=(S_{\hat \theta \hat \phi}, -S_{\hat r \hat \phi} , S_{\hat r \hat \theta}) 
\ .
\eeq
For the CP supplementary conditions only these components of the spin tensor remain nonzero, while in the remaining cases the other nonzero components are determined from these through the corresponding orthogonality condition.
The total spin scalar is also useful
\beq
\label{sinv}
s^2
=\frac12 S_{\mu\nu}S^{\mu\nu}
=-S_{\hat r\hat t }^2 -S_{\hat \theta \hat t }^2 -S_{\hat \phi \hat t }^2
 +S_{\hat r \hat \theta}^2 +S_{\hat r \hat \phi}^2+S_{\hat \theta \hat \phi}^2\ , 
\eeq
and in general is not constant along the trajectory of the spinning particle. In the Schwarzschild field the total spin must be small enough compared to the mass of the test particle and of the black hole $|s|/(mM)\ll 1$ for the approximation of the Mathisson-Papapetrou model to be valid. This inequality follows from requiring that
the characteristic length scale $|s|/m$ associated with the particle's internal structure be  small compared to the natural length scale $M$ associated with the background field in order that the particle backreaction can be neglected, i.e., that the description of a test particle on a background field make sense \cite{mol}.

\section{Solving the equations of motion: preliminary steps}

Consider first the evolution equation for the spin tensor (\ref{papcoreqs2}). 
By contracting both sides of Eq.~(\ref{papcoreqs2}) with $U_\nu$, one obtains the following expression for the total 4-momentum
\begin{equation}
\label{Ps}
P^{\mu}=-(U\cdot P)U^\mu -U_\nu \frac{DS^{\mu\nu}}{\rmd \tau_U}
\equiv
mU^\mu +P_s^\mu\ ,
\end{equation}
which then defines the particle's mass $m$, which a priori does not have to be constant along the orbit, while  $P_s^\mu =U_\alpha DS^{\alpha\mu}/{\rmd \tau_U}$  is the part of the 4-momentum orthogonal to $U$.
Finally let $U_p$ denote the timelike unit vector associated with the total 4-momentum $P=||P||\,U_p\equiv\mu \, U_p$. 

Backsubstituting this representation Eq.~(\ref{Ps}) of the momentum into the spin evolution Eq.~(\ref{papcoreqs2}) expressed in the static observer frame leads to
\begin{eqnarray}
\label{eqs1}
0&=&\frac{\rmd S_{\hat r\hat \phi}}{\rmd \tau_U}-\nu \frac{\rmd S_{\hat t\hat r}}{\rmd \tau_U}+\gamma\frac{\zeta_K}{\nu_K}(\nu^2-\nu_K^2)S_{\hat t\hat \phi}\ , \\
\label{eqs2}
0&=&\frac{\rmd S_{\hat \theta\hat \phi}}{\rmd \tau_U}-\nu \frac{\rmd S_{\hat t\hat \theta}}{\rmd \tau_U}-\frac{\gamma\nu}{\gamma_K^2}\frac{\zeta_K}{\nu_K}S_{\hat r\hat \theta}\ , \\
\label{eqs3}
0&=&\frac{\rmd S_{\hat r\hat \theta}}{\rmd \tau_U}-\gamma\nu_K\zeta_K S_{\hat t\hat \theta}+\gamma\nu\frac{\zeta_K}{\nu_K}S_{\hat \theta\hat \phi}\ .
\end{eqnarray}

From (\ref{Ps}), using the definition of $P_s$ and equations~(\ref{eqs1})--(\ref{eqs3}), it follows that the total 4-momentum $P$ can be written in the form
\begin{eqnarray}\fl\quad
\label{Ptot}
P&=&\gamma(m+\nu m_s)e_{\hat t}
+\frac1{\gamma}\left[\frac{\rmd S_{\hat t\hat r}}{\rmd \tau_U} 
   -\gamma\nu\frac{\zeta_K}{\nu_K} S_{\hat t\hat \phi}\right]e_{\hat r}
+\frac1{\gamma}\left[\frac{\rmd S_{\hat t\hat \theta}}{\rmd \tau_U}
   -\gamma\nu_K\zeta_K S_{\hat r\hat \theta}\right]e_{\hat \theta}
\nonumber\\
\fl\quad
&&+\gamma(m\nu+m_s)e_{\hat \phi}
\nonumber\\
\fl\quad
&=&mU+m_sE_{\hat \phi}+\frac1{\gamma}\left[\frac{\rmd S_{\hat t\hat r}}{\rmd \tau_U}-\gamma\nu\frac{\zeta_K}{\nu_K} S_{\hat t\hat \phi}\right]e_{\hat r}
+\frac1{\gamma}\left[\frac{\rmd S_{\hat t\hat \theta}}{\rmd \tau_U}
       -\gamma\nu_K\zeta_K S_{\hat r\hat \theta}\right]e_{\hat \theta}\ ,
\end{eqnarray}
with
\beq
\label{msdef}
m_s=\frac{\rmd S_{\hat t\hat \phi}}{\rmd \tau_U} 
     +\gamma\nu\frac{\zeta_K}{\nu_K} S_{\hat t\hat r} 
         -\gamma\nu_K\zeta_K S_{\hat r\hat \phi}\ .
\eeq

Next consider the equation of motion (\ref{papcoreqs1}). 
The Riemann tensor spin-curvature-coupling force term is
\begin{eqnarray}\fl\quad
\label{Fspin}
F^{\rm (sc)}
=\gamma\zeta_K^2 \left\{\nu S_{\hat t\hat \phi} e_{\hat t}
+ \left[2S_{\hat t\hat r}+\nu S_{\hat r\hat \phi}\right] e_{\hat r}
-\left[S_{\hat t\hat \theta}+2\nu S_{\hat \theta\hat \phi}\right] e_{\hat \theta}
-S_{\hat t\hat \phi}e_{\hat \phi}\right\}\ .
\end{eqnarray}
Using (\ref{Ps}), the balance condition which allows a circular orbit of this type to exist can be written as
\beq
\label{baleq}
ma(U)=F^{\rm(so)} + F^{\rm (sc)}\ ,
\eeq
where $a(U)$ is the acceleration of the $U$-orbit and $F^{\rm(so)}\equiv -DP_s/d\tau_{U}$ defines the spin-orbit coupling force term, which arises from the variation of the spin along the orbit.

Taking (\ref{Ptot}) and (\ref{msdef}) into account, Eq.~(\ref{papcoreqs1}) gives rise to the following set of ordinary differential equations
\begin{eqnarray}\fl\quad
\label{eqm1}
0&=&\nu\frac{\rmd^2S_{\hat t\hat \phi}}{\rmd \tau_U^2}-2\gamma\nu \nu_K\zeta_K\frac{\rmd S_{\hat r\hat \phi}}{\rmd \tau_U}+\gamma\frac{\zeta_K}{\nu_K}(\nu^2+\nu_K^2)\frac{\rmd S_{\hat t\hat r}}{\rmd \tau_U}\nonumber\\
\fl\quad
&&-\gamma^2\nu\zeta_K^2(\nu^2-\nu_K^2)S_{\hat t\hat \phi}+\frac{\rmd m}{\rmd \tau_U}\ ,  \\
\fl\quad
\label{eqm2}
0&=&\frac{\rmd^2S_{\hat t\hat r}}{\rmd \tau_U^2}-\nu\frac{\rmd^2S_{\hat r\hat \phi}}{\rmd \tau_U^2}-2\frac{\gamma\nu}{\gamma_K^2}\frac{\zeta_K}{\nu_K}\frac{\rmd S_{\hat t\hat \phi}}{\rmd \tau_U}+\nu\gamma^2\zeta_K^2(\nu^2-\nu_K^2)S_{\hat r\hat \phi} \nonumber\\
\fl\quad
&&-\zeta_K^2\left[\frac{\gamma^2}{\gamma_K^2}\frac{\nu^2}{\nu_K^2}+2\right]S_{\hat t\hat r}-m\gamma\frac{\zeta_K}{\nu_K}(\nu^2-\nu_K^2)\ , \\
\fl\quad
\label{eqm3}
0&=&\frac{\rmd^2S_{\hat t\hat \theta}}{\rmd \tau_U^2}-\nu\frac{\rmd^2S_{\hat \theta\hat \phi}}{\rmd \tau_U^2}+\gamma\frac{\zeta_K}{\nu_K}(\nu^2-\nu_K^2)\frac{\rmd S_{\hat r\hat \theta}}{\rmd \tau_U}+2\nu\zeta_K^2S_{\hat \theta\hat \phi}+\zeta_K^2S_{\hat t\hat \theta}\ ,  \\
\fl\quad
\label{eqm4}
0&=&\frac{\rmd^2S_{\hat t\hat \phi}}{\rmd \tau_U^2}-\gamma\frac{\zeta_K}{\nu_K}(\nu^2+\nu_K^2)\frac{\rmd S_{\hat r\hat \phi}}{\rmd \tau_U}+2\gamma\nu\frac{\zeta_K}{\nu_K}\frac{\rmd S_{\hat t\hat r}}{\rmd \tau_U}-\zeta_K^2\left[\frac{\gamma^2}{\gamma_K^2}\frac{\nu^2}{\nu_K^2}-1\right]S_{\hat t\hat \phi}\nonumber\\
\fl\quad
&&+\nu\frac{\rmd m}{\rmd \tau_U}\ . 
\end{eqnarray}
Note that there are two equations containing the second derivative of $S_{\hat t\hat \phi}$; this is due to the presence of its first derivative in two different components of $P$ (more precisely, in $P^{\hat t}$ and $P^{\hat \phi}$, see Eqs.~(\ref{Ptot}) and (\ref{msdef})).

Once the system of constant coefficient linear differential equations (\ref{eqs1})--(\ref{eqs3}) and (\ref{eqm1})--(\ref{eqm4}) is solved for $m$ and the spin tensor components, one may then calculate $P$. The system must be decoupled, leading to functions which are either exponentials, sinusoidals, or at most quadratic functions of the proper time along the particle world line. The elimination method for decoupling the equations is crucially different depending on whether  
$\nu$ has the values 0 or $\pm \nu_K$ or none of these values, since one or the other or neither term drops out of the spin equations (\ref{eqs1}) and so must be considered separately.
From the details of their derivations discussed in the Appendix, one sees why there are several zones approaching the horizon where the solutions change character. 

\section{Particles at rest: the $\nu=0$ case}

When the particle is at rest, the solutions for the components of the spin tensor and the varying mass $m$ of the spinning particle are given by

\begin{enumerate}
\label{solnueq0}

\item $2M<r<3M$:

\begin{eqnarray}
\label{solnueq0I}
S_{\hat \theta\hat \phi}
&=&c_1
\ , \nonumber\\
S_{\hat t\hat r}
&=&c_2e^{\omega_1\tau}+c_3e^{-\omega_1\tau}+\frac{\nu_K}{\zeta_K}\frac{c_m}{2+\nu_K^2}
\ , \nonumber\\
m&=&-\nu_K\zeta_K [c_2e^{\omega_1\tau}+c_3e^{-\omega_1\tau}]+\frac{2c_m}{2+\nu_K^2}
\ , \nonumber\\
S_{\hat t\hat \theta}
&=&c_4 e^{{\bar \omega}_0\tau}+c_5 e^{-{\bar \omega}_0\tau}
\ , \nonumber\\
S_{\hat t\hat \phi}
&=&c_6 e^{{\bar \omega}_0\tau}+c_7 e^{-{\bar \omega}_0\tau}
\ , \nonumber\\
S_{\hat r\hat \theta}
&=&\gamma_K\nu_K\left[c_4 e^{{\bar \omega}_0\tau}-c_5 e^{-{\bar \omega}_0\tau}\right]+c_8
\ , \nonumber\\
S_{\hat r\hat \phi}
&=&\gamma_K\nu_K\left[c_6 e^{{\bar \omega}_0\tau}-c_7 e^{-{\bar \omega}_0\tau}\right]+c_9
\ ;
\end{eqnarray}

\item $r=3M$:

\begin{eqnarray}
\label{solnueq0req3M}
S_{\hat \theta\hat \phi}&=&c_1\ , \nonumber\\
S_{\hat t\hat r}&=&c_2e^{\tau/(3M)}+c_3e^{-\tau/(3M)}+\sqrt{3}Mc_m\ , \nonumber\\
m&=&-\frac{\sqrt{3}}{9M} [c_2e^{\tau/(3M)}+c_3e^{-\tau/(3M)}]+\frac23c_m\ , \nonumber\\
S_{\hat t\hat \theta}&=&c_4\tau+c_5\ , \nonumber\\
S_{\hat t\hat \phi}&=&c_6\tau+c_7\ , \nonumber\\
S_{\hat r\hat \theta}&=&\frac{\sqrt{3}}{9M}\left[\frac{c_4}2\tau^2+c_5\tau\right]+c_8\ , \nonumber\\
S_{\hat r\hat \phi}&=&\frac{\sqrt{3}}{9M}\left[\frac{c_6}2\tau^2+c_7\tau\right]+c_9\ ;
\end{eqnarray}

\item $r>3M$:

\begin{eqnarray}
\label{solnueq0II}
S_{\hat \theta\hat \phi}
&=&c_1
\ , \nonumber\\
S_{\hat t\hat r}
&=&c_2e^{\omega_1\tau}+c_3e^{-\omega_1\tau}+\frac{\nu_K}{\zeta_K}\frac{c_m}{2+\nu_K^2}
\ , \nonumber\\
m&=&-\nu_K\zeta_K [c_2e^{\omega_1\tau}+c_3e^{-\omega_1\tau}]+\frac{2c_m}{2+\nu_K^2}
\ , \nonumber\\
S_{\hat t\hat \theta}
&=&c_4 \cos\omega_0\tau+c_5 \sin\omega_0\tau
\ , \nonumber\\
S_{\hat t\hat \phi}
&=&c_6 \cos\omega_0\tau+c_7 \sin\omega_0\tau
\ , \nonumber\\
S_{\hat r\hat \theta}
&=&\gamma_K\nu_K\left[c_4 \sin\omega_0\tau-c_5 \cos\omega_0\tau\right]+c_8
\ , \nonumber\\
S_{\hat r\hat \phi}
&=&\gamma_K\nu_K\left[c_6 \sin\omega_0\tau-c_7 \cos\omega_0\tau\right]+c_9 
\ ,
\end{eqnarray}

\end{enumerate}
where $c_m, c_1, \ldots, c_9$ are integration constants and
\beq\fl\quad
\label{freqnueq0}
\omega_0=i{\bar \omega}_0=\frac{\zeta_K}{\gamma_K}=\sqrt{\frac{M(r-3M)}{r^3(r-2M)}}
\ , \quad 
\omega_1=\zeta_K(2+\nu_K^2)^{1/2}=\sqrt{\frac{M(2r-3M)}{r^3(r-2M)}}\ .
\eeq

From Eq.~(\ref{Ptot}) the total 4-momentum $P$ then has the value
\beq\fl\quad
P=m e_{\hat t}+\omega_1[c_2e^{\omega_1\tau}-c_3^{-\omega_1\tau}]e_{\hat r}-\frac{\zeta_K}{\nu_K}\left[S_{\hat r\hat \theta}-\frac{c_8}{\gamma_K^2}\right]e_{\hat \theta}-\frac{\zeta_K}{\nu_K}\left[S_{\hat r\hat \phi}-\frac{c_9}{\gamma_K^2}\right]e_{\hat \phi}\ 
\eeq
in cases (i) and (iii), and 
\beq\fl\,
P=m e_{\hat t}+\frac1{3M}[c_2e^{\tau/(3M)}-c_3^{-\tau/(3M)}]e_{\hat r}-\left[\frac{\sqrt{3}}{9M}S_{\hat r\hat \theta}-c_4\right]e_{\hat \theta}-\left[\frac{\sqrt{3}}{9M}S_{\hat r\hat \phi}-c_6\right]e_{\hat \phi}\ 
\eeq
in case (ii).

At this point the supplementary conditions impose constraints on the constants of integration which appear in the solution.
For a particle at rest ($\nu=0$), the CP and P conditions coincide and imply that
$S_{\hat t \hat a}=0$, namely
\beq
c_2=c_3=c_4=c_5=c_6=c_7=0\ , \qquad c_m=0\ ,
\eeq
leaving arbitrary values for $c_1, c_8, c_9$.
As a consequence, $m$ should be $0$ as well, implying that $P$ should be spacelike  and therefore physically inconsistent.

The T supplementary conditions when $\nu=0$ imply instead
\begin{eqnarray}
\label{Tcondsoldnueq0}
0&=&S_{\hat t\hat \theta}\frac{\rmd S_{\hat t\hat \theta }}{\rmd \tau_U}
+S_{\hat t\hat r}\frac{\rmd S_{\hat t\hat r }}{\rmd \tau_U}
+S_{\hat t\hat \phi}\frac{\rmd S_{\hat t\hat \phi }}{\rmd \tau_U}
-\nu_K\zeta_K[S_{\hat t\hat \phi}S_{\hat r\hat \phi}
        +S_{\hat t\hat \theta}S_{\hat r\hat \theta}]
\ , \nonumber\\
0&=&S_{\hat r\hat \theta}\frac{\rmd S_{\hat t\hat \theta }}{\rmd \tau_U}
+S_{\hat r\hat \phi}\frac{\rmd S_{\hat t\hat \phi}}{\rmd \tau_U}
-\nu_K\zeta_K (S_{\hat r\hat \theta}^2+S_{\hat r\hat \phi}^2)
-mS_{\hat t\hat r}
\ , \nonumber\\
0&=&S_{\hat \theta \hat \phi}\frac{\rmd S_{\hat t\hat \phi}}{\rmd \tau_U}
-S_{\hat r\hat \theta}\frac{\rmd S_{\hat t\hat r }}{\rmd \tau_U}
-\nu_K\zeta_K S_{\hat \theta \hat \phi}S_{\hat r\hat \phi}-mS_{\hat t\hat \theta}
\ , \nonumber\\
0&=&S_{\hat \theta\hat \phi}\frac{\rmd S_{\hat t\hat \theta }}{\rmd \tau_U}
+S_{\hat r\hat \phi}\frac{\rmd S_{\hat t\hat r }}{\rmd \tau_U}
-\nu_K\zeta_K S_{\hat r\hat \theta}S_{\hat \theta\hat \phi}
+mS_{\hat t\hat \phi}\ .
\end{eqnarray}
By substituting the solutions given by Eqs.~(\ref{solnueq0I})--(\ref{solnueq0II})  into these equations, one finds that all the integration constants except $c_1$ must vanish.
This in turn implies $m=0$, which again leads to a spacelike $P$.
Thus a spinning particle with nonzero rest mass cannot remain at rest in the given gravitational field.

\section{Geodesic motion: the $\nu=\pm \nu_K$ case}

When the test particle's center of mass moves along a geodesic (the orbit has zero acceleration $a(U)=0$) with azimuthal velocity  $\nu=\pm\nu_K$, the spin-curvature and the spin-orbit forces balance each other (see Eq.~(\ref{baleq})):  $F_{\rm (so)}=-F^{\rm
(sc)}$. The solution of Eqs.~(\ref{eqm1})-(\ref{eqm4}) determines the spin which leads to this balancing.
In the Schwarzschild spacetime, timelike circular geodesics only exist for $r>3M$. We consider
separately the various cases:

\begin{enumerate}
\item $3M<r<6M$:

\begin{eqnarray}\fl\quad
\label{solnuKII}
S_{\hat \theta\hat \phi}&=&c_3\cos\omega_4\tau+c_4\sin\omega_4\tau\ , \nonumber\\
\fl\quad
S_{\hat t\hat \theta}&=&c_5\cos\omega_3\tau+c_6\sin\omega_3\tau\pm\frac{S_{\hat \theta\hat \phi}}{\nu_K}\ , \nonumber\\
\fl\quad
S_{\hat r\hat \theta}&=&\gamma_K\nu_K\left[c_5\sin\omega_3\tau-c_6\cos\omega_3\tau\right]\ , \nonumber\\
\fl\quad
m&=&c_m\ , \nonumber\\
\fl\quad
S_{\hat t\hat r}&=&c_7e^{{\bar \omega}_2\tau}+c_8 e^{-{\bar \omega}_2\tau}\pm2\frac{\gamma_K}{\zeta_K}\frac{c_2}{4-3\gamma_K^2}\ , \nonumber\\
\fl\quad
S_{\hat r\hat \phi}&=&\pm\nu_K[c_7e^{{\bar \omega}_2\tau}+c_8 e^{-{\bar \omega}_2\tau}]+c_1+2\frac{\nu_K}{\zeta_K}\gamma_K\frac{c_2}{4-3\gamma_K^2}\ , \nonumber\\
\fl\quad
S_{\hat t\hat \phi}&=&\pm\frac2{\gamma_K}(3\gamma_K^2-4)^{-1/2}[c_8e^{-{\bar \omega}_2\tau}-c_7e^{{\bar \omega}_2\tau}]+c_9-3\gamma_K^2\frac{c_2}{4-3\gamma_K^2}\tau\ ;
\end{eqnarray}

\item $r=6M$:

\begin{eqnarray}
\label{solnuKreq6M}
S_{\hat \theta\hat \phi}&=&c_3\cos\frac{\sqrt{3}\tau}{18M}+c_4\sin\frac{\sqrt{3}\tau}{18M}\ , \nonumber\\
S_{\hat t\hat \theta}&=&c_5\cos\frac{\sqrt{6}\tau}{36M}+c_6\sin\frac{\sqrt{6}\tau}{36M}\pm2S_{\hat \theta\hat \phi}\ , \nonumber\\
S_{\hat r\hat \theta}&=&\frac{\sqrt{3}}3\left[c_5\sin\frac{\sqrt{6}\tau}{36M}-c_6\cos\frac{\sqrt{6}\tau}{36M}\right]\ , \nonumber\\
m&=&c_m\ , \nonumber\\
S_{\hat t\hat r}&=&c_7\tau+c_8\ , \nonumber\\
S_{\hat r\hat \phi}&=&\pm\frac12[c_7\tau+c_8]+c_1\ , \nonumber\\
S_{\hat t\hat \phi}&=&\mp\frac{\sqrt{2}\tau}{12M}\left[\frac{c_7}2\tau+c_8\right]+c_2\ ;
\end{eqnarray}

\item $r>6M$:

\begin{eqnarray}\fl\quad
\label{solnuKIII}
S_{\hat \theta\hat \phi}&=&c_3\cos\omega_4\tau+c_4\sin\omega_4\tau\ , \nonumber\\
\fl\quad
S_{\hat t\hat \theta}&=&c_5\cos\omega_3\tau+c_6\sin\omega_3\tau\pm\frac{S_{\hat \theta\hat \phi}}{\nu_K}\ , \nonumber\\
\fl\quad
S_{\hat r\hat \theta}&=&\gamma_K\nu_K\left[c_5\sin\omega_3\tau-c_6\cos\omega_3\tau\right]\ , \nonumber\\
\fl\quad
m&=&c_m\ , \nonumber\\
\fl\quad
S_{\hat t\hat r}&=&c_7\cos\omega_2\tau+c_8\sin\omega_2\tau\pm2\frac{\gamma_K}{\zeta_K}\frac{c_2}{4-3\gamma_K^2}\ , \nonumber\\
\fl\quad
S_{\hat r\hat \phi}&=&\pm\nu_K[c_7\cos\omega_2\tau+c_8\sin\omega_2\tau]+c_1+2\frac{\nu_K}{\zeta_K}\gamma_K\frac{c_2}{4-3\gamma_K^2}\ , \nonumber\\
\fl\quad
S_{\hat t\hat \phi}&=&\pm\frac2{\gamma_K}(4-3\gamma_K^2)^{-1/2}[c_8\cos\omega_2\tau-c_7\sin\omega_2\tau]+c_9-3\gamma_K^2\frac{c_2}{4-3\gamma_K^2}\tau\ ,
\end{eqnarray}

\end{enumerate}
where $c_m, c_1, \ldots, c_9$ are integration constants, and three real frequencies are defined for each open interval of radial values by
\begin{eqnarray}\fl\quad
&&\omega_2=i{\bar \omega}_2=\zeta_K(4-3\gamma_K^2)^{1/2}
=\sqrt{\frac{M(r-6M)}{r^3(r-3M)}}\ , \qquad 
\omega_3=\zeta_K=\left(\frac{M}{r^3}\right)^{1/2}\ , \nonumber\\
\fl\quad
&&\omega_4=i{\bar \omega}_4=\zeta_K(3\gamma_K^2-2)^{1/2}
=\frac{1}{r}\sqrt{\frac{M}{r-3M}}
\ .
\end{eqnarray}

Consider first the open interval cases $r\not=6M$.
From Eq.~(\ref{Ptot}), the total 4-momentum $P$ is given by
\begin{eqnarray}\fl\quad
\label{PtotnuK}
P&=&\left\{m\gamma_K\mp\zeta_K\left[S_{\hat r\hat \phi}-(1-2\nu_K^2)\gamma_K^2c_1-2\frac{\nu_K}{\gamma_K\zeta_K}\frac{c_2}{1-4\nu_K^2}\right]\right\}e_{\hat t}\nonumber\\
\fl\quad
&&\mp\frac{\gamma_K^2\zeta_K}{2}\left\{(1+2\nu_K^2)\left[S_{\hat t\hat \phi}+3\frac{c_2}{1-4\nu_K^2}\tau\right]+(1-4\nu_K^2)c_9\right\}e_{\hat r}\nonumber\\
\fl\quad
&&+\left\{ \frac{\zeta_K}{\nu_K}S_{\hat r\hat \theta}\mp(1+2\nu_K^2)^{1/2}\left[c_3 \sin\omega_4\tau -c_4 \cos\omega_4\tau\right] \right\}e_{\hat \theta}\nonumber\\
\fl\quad
&&\pm\left\{m\gamma_K\nu_K\mp\frac{\zeta_K}{\nu_K}\left[S_{\hat r\hat \phi}-(1-2\nu_K^2)\gamma_K^2c_1-2\frac{\nu_K}{\gamma_K}\frac{c_2}{1-4\nu_K^2}\right]\right\}e_{\hat \phi}\ .
\end{eqnarray}

We next impose the standard supplementary conditions.
The CP conditions imply that $S_{\hat t \hat a}=0$, namely
\beq
c_2=c_3=c_4=c_5=c_6=c_7=c_8=c_9=0\ ,
\eeq
so that the only nonvanishing component of the spin tensor is 
$ S^{\hat z}=S_{\hat r\hat \phi}=c_1 \equiv s$, leaving arbitrary values for $c_m$ as well.
From Eq.~(\ref{PtotnuK}), the total 4-momentum $P$ becomes (using $m_s=s\gamma\nu_K\zeta_K$ which follows from Eq.~(\ref{msdef}))
\beq
P=mU_{\pm}+s\gamma\nu_K\zeta_K E_{\hat \phi}\ , 
\eeq
with $U_{\pm}$ given by Eq.~(\ref{Ugeos}). 
Re-examing Eq.~(\ref{Fspin}) shows that the spin-curvature force then acts radially, balancing the radial spin-orbit force.

The P conditions imply 
\beq
S_{\hat t \hat \phi}=0\ , \qquad S_{\hat r \hat t}\pm\nu_K S_{\hat r \hat \phi}=0\ , \qquad S_{\hat \theta  \hat t}\pm\nu_K S_{\hat \theta  \hat \phi}=0\ , 
\eeq
which lead only to the trivial solution 
\beq
c_1=c_2=c_3=c_4=c_5=c_6=c_7=c_8=c_9=0\ ,
\eeq
with $c_m$ arbitrary, or in other words the components of the spin tensor must all be zero, which means that a non-zero spin is incompatible with geodesic motion for a spinning particle. 

The T supplementary conditions when $\nu=\pm\nu_K$ imply
\begin{eqnarray}\fl\quad
\label{TcondsoldnuK}
0&=&S_{\hat t\hat \theta}\frac{\rmd S_{\hat t\hat \theta }}{\rmd \tau_U}+S_{\hat t\hat r}\frac{\rmd S_{\hat t\hat r }}{\rmd \tau_U}+\gamma_K^2S_{\hat t\hat \phi}\frac{\rmd S_{\hat t\hat \phi }}{\rmd \tau_U}\pm m\gamma_K^2\nu_K S_{\hat t\hat \phi}\nonumber\\
\fl\quad
&&\mp\gamma_K\zeta_K\{[S_{\hat t\hat r}S_{\hat t\hat \phi}\pm\nu_K S_{\hat t\hat \theta}S_{\hat r\hat \phi}]-\gamma_K^2S_{\hat t\hat \phi}[S_{\hat t\hat r}\mp\nu_K S_{\hat r\hat \phi}]\}\ , \nonumber\\
\fl\quad
0&=&S_{\hat r\hat \theta}\frac{\rmd S_{\hat t\hat \theta }}{\rmd \tau_U}-\gamma_K^2[\pm\nu_K S_{\hat t\hat r}-S_{\hat r\hat \phi}]\frac{\rmd S_{\hat t\hat \phi}}{\rmd \tau_U}+\gamma_K\frac{\zeta_K}{\nu_K}[S_{\hat t\hat r}^2-\nu_K^2 S_{\hat r\hat \theta}^2]\nonumber\\
\fl\quad
&&-\gamma_K^3\frac{\zeta_K}{\nu_K}\left[S_{\hat t\hat r}^2\mp\nu_K(1+\nu_K^2)S_{\hat t\hat r}S_{\hat r\hat \phi}+\nu_K^2S_{\hat r\hat \phi}^2\right]-m\gamma_K^2[S_{\hat t\hat r}\mp\nu_K S_{\hat r\hat \phi}]\ , \nonumber\\
\fl\quad
0&=&\gamma_K^2[\pm\nu_K S_{\hat t\hat \theta}-S_{\hat \theta \hat \phi}]\left[\frac{\rmd S_{\hat t\hat \phi}}{\rmd \tau_U}-\gamma_K\nu_K\zeta_K S_{\hat r\hat \phi}\right]+S_{\hat r\hat \theta}\frac{\rmd S_{\hat t\hat r }}{\rmd \tau_U}\nonumber\\
\fl\quad
&&+\gamma_K^2[S_{\hat t\hat \theta}\mp\nu_K S_{\hat \theta \hat \phi}]\left[m+\gamma_K\frac{\zeta_K}{\nu_K}S_{\hat t\hat r}\right]-\gamma_K\frac{\zeta_K}{\nu_K}[S_{\hat t\hat r}S_{\hat t\hat \theta}\pm\nu_K S_{\hat r\hat \theta}S_{\hat t\hat \phi}]\ , \nonumber\\
\fl\quad
0&=&\pm\gamma_K^2\nu_K S_{\hat t\hat \phi}\frac{\rmd S_{\hat t\hat \phi }}{\rmd \tau_U}+S_{\hat \theta\hat \phi}\frac{\rmd S_{\hat t\hat \theta }}{\rmd \tau_U}+S_{\hat r\hat \phi}\frac{\rmd S_{\hat t\hat r }}{\rmd \tau_U}+\gamma_K^3\frac{\zeta_K}{\nu_K}S_{\hat t\hat \phi}[S_{\hat t\hat r}\mp\nu_K^3S_{\hat r\hat \phi}]\nonumber\\
\fl\quad
&&+m\gamma_K^2S_{\hat t\hat \phi}-\gamma_K\frac{\zeta_K}{\nu_K}[\nu_K^2S_{\hat r\hat \theta}S_{\hat \theta\hat \phi}+S_{\hat t\hat \phi}(S_{\hat t\hat r}\pm\nu_K S_{\hat r\hat \phi})]\ .
\end{eqnarray}
By substituting into these equations the solutions given by Eqs.~(\ref{solnuKII})--(\ref{solnuKIII}), we obtain the conditions
\beq
c_2=c_3=c_4=c_5=c_6=c_7=c_8=0\ ,
\eeq
implying that the only nonvanishing components of the spin tensor are
\beq
S^{\hat z} =-S_{\hat r \hat \phi}=-c_1\ , \qquad S_{\hat t \hat \phi}=c_9\ , 
\eeq
and either
\beq
c_1, c_9\quad \mbox{arbitrary}\ , \qquad c_m=\pm\gamma_K\zeta_K c_1\ ,
\eeq
which implies that spin component $S^{\hat z}$ is proportional to the mass, locking them together by a constant of proportionality depending on the orbit velocity, or
\beq
\label{physcond1}
c_1=0=c_9\ , \qquad c_m \quad \mbox{arbitrary}\ ,
\eeq
corresponding to the zero spin case where geodesic motion is of course allowed.
In the former case the total spin invariant (\ref{sinv}) reduces to
\beq
s^2=-c_9^2+c_1^2\ ,
\eeq
so that the condition $|s|/(mM)\ll1$ preserving the validity of the Mathisson-Papapetrou model reads
\beq
\frac{|s|}{mM}=\frac{1}{M\gamma_K\zeta_K}\left(1-\frac{c_9^2}{c_1^2}\right)^{1/2}\ll1\ ,
\eeq
implying either $c_1\gtrsim c_9$ or $r\simeq 3M$ (where $\gamma_K\to\infty$).
In the limit $r\to3M$ where the circular geodesics become null and require a separate treatment, one has a solution for which this spin component $S^{\hat z}$ is fixed to have a value 
determined by the constant mass $m$ and the azimuthal velocity, the $t$-$\phi$ component of the spin is arbitrary. If one takes the limit $m\to0$, then the component of the spin vector out of the orbit vanishes, leaving the spin vector locked to the direction of motion as found
by \cite{mashnull} who discussed the null geodesic case using the P supplementary conditions, the latter being the only physically relevant in such limit. 

Finally consider the remaining case $r=6M$.
Eq.~(\ref{Ptot}) then shows that the total 4-momentum $P$ is given by
\begin{eqnarray}\fl\quad
\label{PtotnuKreq6M}
P&=&\left\{\frac23\sqrt{3}m\mp\frac{\sqrt{6}}{108M}\left[3S_{\hat r\hat \phi}-2c_1\right]\right\}e_{\hat t}+\left\{\mp\frac{\sqrt{6}}{36M}S_{\hat t\hat \phi}+\frac{\sqrt{3}}2c_7\right\}e_{\hat r}\nonumber\\
\fl\quad
&&-\left\{\frac{\sqrt{6}}{18M}S_{\hat r\hat \theta}\pm\frac{1}{6M}\left[c_3 \sin\frac{\sqrt{3}\tau}{18M} -c_4 \cos\frac{\sqrt{3}\tau}{18M}\right]\right\}e_{\hat \theta}\nonumber\\
\fl\quad
&&\pm\frac12\left\{\frac23\sqrt{3}m\mp\frac{\sqrt{6}}{108M}\left[3S_{\hat r\hat \phi}-2c_1\right]\right\}e_{\hat \phi}\ .
\end{eqnarray}
Imposing the standard supplementary conditions gives rise to the same result as for the general case $r\not=6M$.

The CP conditions imply 
\beq
c_2=c_3=c_4=c_5=c_6=c_7=c_8=0\ ,
\eeq
so that the only nonvanishing component of the spin tensor is $S^{\hat z}=-S_{\hat r\hat \phi}=-c_1$, for arbitrary values of $c_m$,  leading to constant mass $m$. 
The P conditions give only the trivial solution 
\beq
c_1=c_2=c_3=c_4=c_5=c_6=c_7=c_8=0\ ,
\eeq
with $c_m$ arbitrary, leading to constant mass $m$. 
Finally, the T conditions imply
\beq
c_3=c_4=c_5=c_6=c_7=c_8=0\ ,
\eeq
so that the only nonvanishing components of the spin tensor are
\beq
S^{\hat z} =-S_{\hat r \hat \phi}=-c_1\ , \qquad S_{\hat t \hat \phi}=c_2\ , 
\eeq
and either 
\beq
c_1, c_2\quad \mbox{arbitrary}\ , \qquad c_m=\pm\frac{\sqrt{2}}{18M}c_1\ 
\eeq
or
\beq
\label{physcond2}
c_1=0=c_2\ , \qquad c_m \quad \mbox{arbitrary}\ ,
\eeq
with constant mass $m$ in both cases. 
In the former case the spin invariant (\ref{sinv}) reduces to
\beq
s^2=-c_2^2+c_1^2\ ,
\eeq
so that the condition $|s|/(mM)\ll1$ reads
\beq
\frac{|s|}{mM}=\frac{18}{\sqrt{2}}\left(1-\frac{c_2^2}{c_1^2}\right)^{1/2}\ll1\ ,
\eeq
implying $c_1\gtrsim c_2$.

Thus if the center of mass of the test particle is constrained to be a circular geodesic, either the spin is forced to be zero or have an arbitrary constant value of the single nonzero component $S^{\hat z}$ of the spin vector out of the plane of the orbit.

\section{The general case: $\nu\not=0$ and $\nu\not= \pm \nu_K$}

\begin{figure} 
\typeout{*** EPS figure 1}
\begin{center}
\includegraphics[scale=0.35]{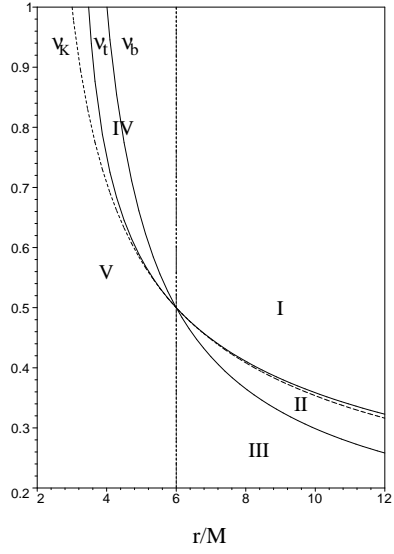}
\end{center}
\caption{The azimuthal velocities ${\bar \nu}$ ($\nu_{\rm b}$ in the plot), ${\tilde \nu}$ ($\nu_{\rm t}$ in the plot) and $\nu_K$ (dashed curve) are plotted as functions of the radial parameter $r/M$. The dashed vertical line corresponds to the value $r/M=6$, where they all coincide. There are five different regions (explicitly indicated in the plot), depending on the relative sign of the azimuthal velocity $\nu$ with respect to ${\bar \nu}$ and ${\tilde \nu}$, which correspond to intervals for which $\Omega_{\pm}$ are real, purely imaginary or neither, as explained in the text.
}  
\label{fig:1}
\end{figure}

For general circular orbits excluding the previous cases $\nu=0$ and $\nu= \pm \nu_K$, the solutions of the equations of motion for the components of the spin tensor and the mass $m$ of the spinning particle are
\begin{eqnarray}\fl\quad
\label{sthphisol}
S_{\hat \theta\hat \phi}&=&A\cos\Omega\tau+B\sin\Omega\tau\ , \\
\fl\quad
\label{stthsol}
S_{\hat t\hat \theta}&=&C\cos\Omega_1\tau+D\sin\Omega_1\tau
+F\, S_{\hat \theta\hat \phi}\ ,\quad 
F=\frac{3\nu \nu_K^2}{[\nu^2(1+2\nu_K^2)-\nu_K^2(1-\nu_K^2)]}\ , \\
\fl\quad
\label{srthsol}
S_{\hat r\hat \theta}
&=&-\frac{\gamma\nu\zeta_K(1+2\nu_K^2)(\nu^2-\nu_K^2)}{\Omega\nu_K[\nu^2(1+2\nu_K^2)
-\nu_K^2(1-\nu_K^2)]}[A\sin\Omega\tau-B\cos\Omega\tau]
\nonumber\\
\fl\quad
&&-\nu_K\gamma_K[D\cos\Omega_1\tau-C\sin\Omega_1\tau]\ , \\
\fl\quad
\label{eqm4e}
S_{\hat r\hat \phi}
&=&\frac{\nu_K}{\zeta_K}\frac{\nu^2-\nu_K^2}{\frac{\nu^2}{\gamma_K^2}+\nu_K^2(2+\nu_K^2)} \left[\gamma\nu c_m - \gamma_K^2\frac{\nu_K}{\zeta_K} \frac{\nu^2(1-4\nu_K^2) +\nu_K^2(2+\nu_K^2)}{(\nu^2-\nu_K^2)^2}c_0\right]
\nonumber\\
\fl\quad
&&+c_1 e^{\Omega_+\tau}+c_2 e^{-\Omega_+\tau}+c_3 e^{\Omega_-\tau}+c_4 e^{-\Omega_-\tau}
\ , \\
\fl\quad
\label{eqm2d}
S_{\hat t\hat r}
&=&\frac{\nu_K}{\zeta_K}\frac{1}{\frac{\nu^2}{\gamma_K^2}+\nu_K^2(2+\nu_K^2)} \left[-\gamma (\nu^2-\nu_K^2)c_m+\nu\frac{\nu_K}{\zeta_K}c_0\right] +\frac{1}{2\nu\nu_K^2(1+2\nu_K^2)}\cdot\nonumber\\
\fl\quad
&&\cdot\{(3\nu_K^2+\Phi)
\left[c_1 e^{\Omega_+\tau} +c_2 e^{-\Omega_+\tau}\right] 
+(3\nu_K^2-\Phi)\left[c_3 e^{\Omega_-\tau}+c_4 e^{-\Omega_-\tau}\right]\}\ , \\
\fl\quad
\label{eqs1c}
S_{\hat t\hat \phi}
&=&\frac{1}{\gamma}\frac{\nu_K}{\zeta_K}\frac1{\nu^2-\nu_K^2} 
\bigg\{\Omega_+\left[\frac{3\nu_K^2+\Phi}{2\nu_K^2(1+2\nu_K^2)}-1\right] 
\left[c_1 e^{\Omega_+\tau}-c_2 e^{-\Omega_+\tau}\right]
\nonumber\\
\fl\quad
&&+\Omega_-\left[\frac{3\nu_K^2-\Phi}{2\nu_K^2(1+2\nu_K^2)}-1\right] \left[c_3 e^{\Omega_-\tau}-c_4 e^{-\Omega_-\tau}\right]\bigg\}\ , \\ 
\fl\quad
\label{massvar}
m&=&\gamma\frac{\zeta_K}{\nu_K}[\nu S_{\hat r\hat \phi}- \nu_K^2 S_{\hat t\hat r}]+c_m\ ;
\end{eqnarray}
where $A, B, C, D, c_m, c_0, \ldots, c_4$ are integration constants, and the real positive frequencies $\Omega$ and $\Omega_1$ are given by
\beq
\Omega=\gamma\zeta_K (1+2\nu_K^2)^{1/2}\frac{|\nu|}{\nu_K} \ , \qquad 
\Omega_1=\frac{\gamma\zeta_K}{\gamma_K}\ ,
\eeq
assumed to be distinct for the above equations to be valid,
and the remaining abbreviations are
\beq
\label{Omegapm}
\Omega_{\pm}=-\frac{\gamma}{\gamma_K}\frac{\zeta_K}{\nu_K}\left[{\bar \nu}^2-\nu^2\pm\frac{\gamma_K^2}2\Phi\right]^{1/2}, \quad
\Phi=3\nu_K^2\left[1-\frac{\nu^2}{{\tilde \nu}^2}\right]^{1/2},
\eeq
with
\beq
{\bar \nu}^2=\frac{\gamma_K^2\nu_K^2}{2}(1+2\nu_K^2)\ , \quad 
{\tilde \nu}^2=\frac98\frac{\gamma_K^2\nu_K^2}{(1+2\nu_K^2)}\ .
\eeq
The behaviors of the azimuthal velocities ${\bar \nu}$, ${\tilde \nu}$ and $\nu_K$ as functions of the radial parameter $r/M$ are compared in Fig.~\ref{fig:1}. They all coincide at $r=6M$, where ${\bar \nu}={\tilde \nu}=\nu_K=1/2$; for $2M<r<6M$ it is ${\tilde \nu}<{\bar \nu}$, while ${\tilde \nu}>{\bar \nu}$ for $r>6M$.


The quantities $\Omega_{\pm}$ also lead to angular velocities for certain intervals of values of the azimuthal velocity $\nu$. In fact we are interested in those values for which $\Omega_{+}$ and/or $\Omega_{-}$ are purely imaginary, since the imaginary parts can be interpreted as additional frequencies characterizing spin precession. 
One must distinguish the cases $2M<r<6M$ and $r>6M$, referring to Fig.~\ref{fig:1} and to Eq.~(\ref{Omegapm}):

\begin{itemize}

\item[a)] $r>6M$:

\begin{itemize}
\item[-] if $\nu>{\tilde \nu}$ (region I), the quantities $\Omega_{\pm}$ are both complex; 
\item[-] if $\nu={\tilde \nu}$, $\Omega_{+}=\Omega_{-}$ is purely imaginary, since ${\tilde \nu}>{\bar \nu}$;
\item[-] if ${\bar \nu}<\nu<{\tilde \nu}$ (region II), $\Omega_{-}$ is purely imaginary, while $\Omega_{+}$ can be either real (even zero) or purely imaginary;
\item[-] if $\nu<{\bar \nu}$ (region III), $\Omega_{+}$ is purely imaginary, while $\Omega_{-}$ can be either real (even zero) or purely imaginary;
\end{itemize}

\item[b)] $2M<r<6M$:

\begin{itemize}
\item[-] if $\nu>{\tilde \nu}$ (region IV), the quantities $\Omega_{\pm}$ are both complex; 
\item[-] if $\nu={\tilde \nu}$, $\Omega_{+}=\Omega_{-}$ is real, since ${\tilde \nu}<{\bar \nu}$;
\item[-] if $\nu<{\tilde \nu}$ (region V), $\Omega_{+}$ is real, since ${\tilde \nu}<{\bar \nu}$, while $\Omega_{-}$ can be either real (even zero) or purely imaginary.
\end{itemize}

\end{itemize}

All of these remarks so far assume that the two frequencies $\Omega$ and $\Omega_1$ are distinct, necessary for the decoupling procedure which leads to this solution. A different result follows in the special case $\Omega=\Omega_1$. This occurs for the particular value of the azimuthal velocity 
\beq
\label{nusolomeqom1}
\nu_0=\pm\frac{\nu_K}{\gamma_K}(1+2\nu_K^2)^{-1/2}
=\pm \sqrt{\frac{M(r-3M)}{r(r-2M)}}\ ,
\eeq
which vanishes at $r=3M$ and is real for $r>3M$,
rising to its peak speed at $r\approx 3.934M$ and decreasing asymptotically towards 
the geodesic speed from below as $r\to \infty$.
The solutions for the components $S_{\hat \theta\hat \phi}$, $S_{\hat t\hat \theta}$ and $S_{\hat r\hat \theta}$ of the spin tensor are given by
\begin{eqnarray}\fl\quad
\label{sthphisolomeqom1}
S_{\hat \theta\hat \phi}&=&A\cos\Omega\tau+B\sin\Omega\tau\ , \\
\fl\quad
\label{stthsolomeqom1}
S_{\hat t\hat \theta}&=&\left[C-\frac32\frac{\gamma_0^2\nu_0}{\Omega}\zeta_K^2(A-B\Omega\tau)\right]\cos\Omega\tau+\left[D-\frac32\frac{\gamma_0^2\nu_0}{\Omega^2}\zeta_K^2A\tau\right]\sin\Omega\tau\ , \\
\fl\quad
\label{srthsolomeqom1}
S_{\hat r\hat \theta}
&=&\frac12\frac{\zeta_K}{\nu_K}\frac{\gamma_0^3}{\Omega^2}\bigg\{
\left[3\nu_0\nu_K^2\zeta_K^2(B+A\Omega\tau)+2\frac{\Omega^2}{\gamma_0^2}(\nu_0 B-\nu_K^2D)\right]\cos\Omega\tau \nonumber\\
\fl\quad
&&+\left[3\nu_0\nu_K^2\zeta_K^2(2A+B\Omega\tau)-2\frac{\Omega^2}{\gamma_0^2}(\nu_0 A-\nu_K^2C)\right]\sin\Omega\tau
\bigg\}\ ,
\end{eqnarray}
with
\beq\fl\quad
\Omega\equiv\Omega_1=\frac{\zeta_K}{\nu_K}\left[\frac{1+2\nu_K^2}{1+\nu_K^2(1+\nu_K^2)}\right]^{1/2}=\frac{\sqrt{M}}{r}\left[\frac{r-3M}{r(r-3M)+3M^2}\right]^{1/2}\ ,
\eeq
while that corresponding to the remaining components as well as to the varying mass $m$ are obtained simply by evaluating the general solutions (\ref{eqm4e})--(\ref{massvar}) at $\nu=\nu_0$. 
The reality properties of the quantities $\Omega_{\pm}$ can be determined as done in the general case, noting that  $\nu_0<{\tilde \nu}$ (corresponding to region V) always holds in the interval $3M<r<6M$. For $r>6M$ however, we must distinguish two different regions, 
a) $6M<r<{\bar r}_0$, with ${\bar r}_0=6M(1+\sqrt{2}/2)\approx10.24M$ such that $\nu_0={\bar \nu}$, where $\nu_0<{\bar \nu}$ (corresponding to region III), and 
b) $r>{\bar r}_0$, where $\nu_0>{\bar \nu}$ (corresponding to region II).

The behavior of $S$, $U$ and $P$ along the world line itself is completely 
determined by the initial conditions
\begin{eqnarray}
&& S_{\hat \alpha\hat \beta}(0)\ ,\quad 
\frac{\rmd S_{\hat \alpha\hat \beta}}{\rmd \tau_U}\bigg\vert_{\tau=0}\ ,
\end{eqnarray}
and the corresponding conditions on the mass $m$ of the particle which follow from Eq.~(\ref{massvar}).
Thus in the special case in which the ``center of mass line" is directed along a circular orbit, the completion of the scheme for the spinning test particle is equivalent to a choice of initial conditions.  

In principle the components of the spin tensor which are not constants should precess with the 
different frequencies which appear in 
Eqs.~(\ref{sthphisol})--(\ref{eqs1c}), leading to non-periodic motion, a feature that seems to characterize the  general situation in the Schwarzschild \cite{maeda} and Kerr \cite{semerak,hartl1,hartl2} spacetimes.
However, this does not occur in practice once the CP, P and T supplementary conditions are imposed, as we will
see below. It turns out that the nonvanishing components of the spin tensor are all constant in the CP and T cases,
while the motion is periodic with a unique frequency in the P case. As one might expect, the particle mass $m$ turns out to be constant in all three cases. 

\subsection{The CP supplementary conditions}

The CP supplementary conditions require 
\beq
S_{\hat t \hat r}=0\ , \qquad S_{\hat t \hat \theta}=0\ , \qquad S_{\hat t \hat \phi}=0\ . 
\eeq
From Eq.~(\ref{eqm2d}) this forces 
\beq
c_1=c_2=c_3=c_4=0\ , \qquad 
c_0=\frac{\gamma}{\nu}\frac{\zeta_K}{\nu_K}(\nu^2-\nu_K^2)c_m\ . 
\eeq
Substituting these values into Eq.~(\ref{eqm4e}) then gives 
\beq
c_m=-\frac{\gamma\nu}{\gamma_K^2}\frac{\zeta_K}{\nu_K}S_{\hat r \hat \phi}\ , 
\eeq
so from Eq.~(\ref{massvar}) we get 
\beq
\label{ssolCPgen}
S_{\hat r \hat \phi}=s=\frac{m}{\gamma\nu}\frac1{\nu_K\zeta_K}\ . 
\eeq
Finally, from Eqs.~(\ref{sthphisol}) and (\ref{srthsol}) it follows that
\beq
S_{\hat \theta \hat \phi}=0\ , \qquad S_{\hat r \hat \theta}=0\ . 
\eeq
However, from Eq.~(\ref{Ptot}) it follows that
\beq
P=-s\nu_K\zeta_K e_{\hat \phi}\ , 
\eeq
since $m_s=-s\gamma\nu_K\zeta_K=-m/\nu$, a consequence of Eqs.~(\ref{msdef}) and (\ref{ssolCPgen}).
This result is unphysical since the total 4-momentum $P$ is spacelike. 

\subsection{The P supplementary conditions}

The P supplementary conditions require 
\beq
\label{Pconds}
S_{\hat t \hat \phi}=0\ , \qquad 
S_{\hat r \hat t}+S_{\hat r \hat \phi}\nu=0\ , \qquad 
S_{\hat \theta  \hat t}+S_{\hat \theta  \hat \phi}\nu=0\ . 
\eeq
Under these conditions the components of the spin vector $S_U$ in the local rest space of the particle, 
$S_U^\beta=\frac12 \eta_\alpha{}^{\beta\gamma\delta}U^\alpha S_{\gamma\delta}$,
expressed in the Frenet-Serret frame, are just
\beq
\label{spinFS}
(S_U^{1},S_U^{2},S_U^{3})
=(\gamma^{-1} S_{\hat \theta \hat \phi}, 
S_{\hat r \hat \theta}, 
\gamma^{-1} S_{\hat r \hat \phi}) 
\ .
\eeq

Comparing the first Eq.~(\ref{Pconds}) with Eq.~(\ref{eqs1c}) we get 
\beq
c_1=c_2=c_3=c_4=0\ , 
\eeq
so that $S_{\hat t \hat r}$, $S_{\hat r \hat \phi}$ and the particle mass $m$ are all constant.
Eqs.~(\ref{eqm4e}) and  (\ref{eqm2d}) together with the second of the Pirani conditions Eq.~(\ref{Pconds}) imply
\beq
c_0=\frac{\gamma}{\nu}\frac{\zeta_K}{\nu_K}\frac{1+\nu^2}{\gamma_K^2}\frac{(\nu^2-\nu_K^2)^2}{\nu^2(2-5\nu_K^2)+\nu_K^2(1+2\nu_K^2)}\,c_m\ ,
\eeq
hence from Eq.~(\ref{massvar})
\beq
c_m=\left[1+\frac{1}{\gamma_K^2}\frac{\nu^2-\nu_K^2}{\nu^2(1-4\nu_K^2)+\nu_K^2(2+\nu_K^2)}\right]m\ .
\eeq
Next by substituting these values of the constants $c_0$ and $c_m$ into Eq.~(\ref{eqm4e}), we obtain
\beq\label{eq:S3}
\gamma S_U^3 =S^{\hat z}
=S_{\hat r \hat \phi}=-m\frac{\gamma}{\nu}\frac{\nu_K}{\zeta_K}\frac{\nu^2-\nu_K^2}{\frac{\gamma^2}{\gamma_K^2}(\nu^2-\nu_K^2)+3\nu_K^2}\ .
\eeq

Finally comparing the last of the Pirani conditions Eq.~(\ref{Pconds}) with  Eqs.~(\ref{sthphisol})--(\ref{stthsol}) and Eqs.~(\ref{sthphisolomeqom1})--(\ref{stthsolomeqom1}) leads to two possibilities: either
\beq
\hbox{case P1:\qquad}
A=B=C=D=0\ ,
\eeq
which places no constraint on $\nu$ and the spin vector is constant and out of the plane of the orbit, or 
\beq
\hbox{case P2:\qquad} 
C=0=D\ , \quad F=\nu\ ,
\eeq
the latter of which (again from Eq.~(\ref{stthsol})) leads to the special azimuthal velocity
\beq\fl\quad
\label{nupirani2}
\nu = \nu_{(P2)}
=\pm 2\nu_K \left(\frac{1-\nu_K^2/4}{1+2\nu_K^2}\right)^{1/2}
=\pm 2\left(\frac{M(r-9M/4)}{r(r-2M)}\right)^{1/2}\ .
\eeq
The case P1 has been already considered previously \cite{bdfg1},
leaving only P2 to be considered here. 
The corresponding azimuthal speed $\nu_{(P2)}$ vanishes at $r=9/4M$ and is real for $r>9/4M$,
rising to a maximum speed of 1 at $r=3M$, corresponding to the two null circular geodesics, and 
decreasing asymptotically towards 
twice the geodesic speed from below as $r\to \infty$.

The corresponding values of $\gamma$ and $\Omega$ are respectively
\beq
\label{eq:gammaP2}
\gamma_{(P2)}=\frac{(1+2\nu_K^2)^{1/2}}{1-\nu_K^2}=\frac{\sqrt{r(r-2M)}}{r-3M}\ , 
\eeq
and
\beq
\label{eq:OmegaP2}
\Omega_{(P2)}
= \gamma_K^2\zeta_K[(4-\nu_K^2)(1+2\nu_K^2)]^{1/2}
= \frac{\sqrt{M}}{r}\frac{\sqrt{4r-9M}}{r-3M}\ , 
\eeq
using Eq.~(\ref{eq:phitau}).
To get the anglular velocity of precession with respect to a frame which is nonrotating with respect to infinity, one must subtract away the precession angular velocity $\Omega_U=\gamma\nu/r$ of the spherical frame. In the case P2 one finds
$\Omega_{(P2)}-\Omega_U=0$, so the spin does not precess with respect to a frame which is nonrotating at infinity.

Substituting these values back into Eq.~(\ref{eq:S3}) then leads to
\beq
S_U^{3}
= \pm\frac{m}{2\Omega_{(P2)}}\ .
\eeq
The remaining nonzero spin components (\ref{sthphisol})--(\ref{srthsol}) can then be expressed in the form
\begin{eqnarray}
\label{spinsolsP}
S_U^1=\gamma^{-1}S_{\hat \theta \hat \phi}
&=&\gamma^{-1}[ A\cos\Omega_{(P2)}\tau_U+B\sin\Omega_{(P2)}\tau_U]\ ,\nonumber\\
S_U^2=S_{\hat r \hat \theta}
&=&\gamma^{-1} [A\sin\Omega_{(P2)}\tau_U-B\cos\Omega_{(P2)}\tau_U]\ , 
\end{eqnarray}

\begin{eqnarray}
\fl\quad
S_U^1(0)=\gamma^{-1}A\ ,  \qquad
S_U^2(0)=-\gamma^{-1}B\ ,  \qquad 
S_U^3(0)=\pm\frac{m}{2\Omega_{(P2)}}\ , 
\end{eqnarray}
leading to
\beq
\pmatrix{
S_U^{1}\cr
S_U^{2}\cr
S_U^{3}\cr}=
\pmatrix{\cos\phi  &\sin\phi & 0\cr
	-\sin\phi  &\cos\phi & 0\cr
         	0 & 0& 1\cr}
\pmatrix{
S_U^{1}(0)\cr
S_U^{2}(0)\cr
S_U^{3}(0)\cr}\ .
\eeq

The spin invariant (\ref{sinv}) becomes in this case
\beq
s^2=\frac{1}{\gamma^2}\left[A^2+B^2+\frac{m^2}{4\zeta_K^2}\frac1{4-\nu_K^2}\right]\ .
\eeq
The Mathisson-Papapetrou model is valid  if the condition $|s|/(mM)\ll1$ is satisfied. 
From the previous equation we have that either $\gamma\to\infty$ or the sum of the bracketed terms must be small, i.e., 
\beq
\label{scoeffs}
\frac{A^2}{m^2M^2}\ll1\ , \quad \frac{B^2}{m^2M^2}\ll1\ , \quad [4M^2\zeta_K^2(4-\nu_K^2)]^{-1}\ll1\ .
\eeq
The latter possibility cannot occur for any allowed values of $r/M$, since the third term (which is dimensionless) of (\ref{scoeffs}) is always greater than $\approx1.88$, as is easily verified. The former possibility is realized only in the case of ultrarelativistic motion, which Eq.~(\ref{eq:gammaP2}) implies occurs only as $r\to 3M$, where the orbits approach null geodesics and the limit $m\to0$ forces the component of the spin vector out of the plane of the orbit to vanish, locking the spin vector to the direction of motion exactly as discussed by Mashhoon \cite{mashnull}.

It is well known that the spin vector $S_U = S_U^i E_i$ lying in the local rest space of $U$ is Fermi-Walker transported along $U$ in the P case, so it must satisfy
\beq\fl\quad
\label{fweq}
0=\frac{D_{(\rm fw)}S_U}{\rmd \tau_U}\equiv P(U)\frac{DS_U}{\rmd \tau_U}
=
 \left[\frac{\rmd S_U^{1}}{\rmd \tau_U}+S_U^2\tau_1 \right] E_1
+\left[\frac{\rmd S_U^{2}}{\rmd \tau_U}-S_U^1\tau_1 \right] E_2\ ,
\eeq
from (\ref{FSeqs}), where $P(U)^\mu_\alpha=\delta^\mu_\alpha+U^\mu U_\alpha$  projects into the local rest space of $U$.   
To check this we must show that the following two equations are identically satisfied
\beq
\label{fwconds}
\frac{\rmd S_U^{1}}{\rmd \tau_U}+\tau_1S_U^2=0\ , \qquad 
\frac{\rmd S_U^{2}}{\rmd \tau_U}-\tau_1 S_U^1=0\ .
\eeq
But these two equations follow immediately from (\ref{spinsolsP}), since $\tau_1=\Omega_{(P2)}$ results from the direct evaluation of the expression (\ref{kappatau1}) for  $\tau_1$, with $\nu$ given by (\ref{nupirani2}).

Thus given the rest mass $m$ of the test particle, the constant component of the spin orthogonal to the orbit is then fixed by the orbit parameters, while the component in the plane of the orbit as seen within the local rest space of the particle itself is locked to a direction which is fixed with respect to the distant observers,
since the angle of precession with respect to the spherical axes is exactly the azimuthal angle of the orbit, but in the opposite sense. In other words the precession of the spin, which introduces a time varying force into the mix, must be locked to the first torsion of the orbit itself in order to maintain the alignment of the 4-velocity with a static direction in the spacetime, and the spin does not precess with respect to observers at spatial infinity. 
Furthermore, the specific spin of the test particle cannot be made arbitrarily small except near the limiting radius where the 4-velocity of this solution goes null, and the spin vector is then locked to the direction of motion.
Apparently the imposition of a circular orbit on the center of mass world line of the test particle is just too strong a condition to describe an interesting spin precession.

The total 4-momentum $P$ given by Eqs.~(\ref{Ptot}) and  (\ref{msdef}) can be written in this case as
\begin{eqnarray}
\label{PtotP}
P &=& mU + m_s E_{2}+\gamma^{-1}\left[\frac{\rmd S_U^{2}}{\rmd \tau_U}
  -\gamma\nu_K\zeta_K S_{\hat r\hat \theta}\right]e_{\hat \theta}
\nonumber\\
&=& mU+m_s E_{2}-\left(\frac{\gamma\nu^2}r +\nu_K \zeta_K\right)S_U^2 e_{\hat z}
\ ,
\end{eqnarray}
with $\nu$ given by (\ref{nupirani2}) and $m_s$ a constant
\beq
\label{msP}
m_s=\gamma\frac{\zeta_K}{\nu_K}(\nu S_{\hat t\hat r}-\nu_K^2 S_{\hat r\hat \phi})
=\gamma^2\frac{\zeta_K}{\nu_K}(\nu^2-\nu_K^2)S_U^3\ ,
\eeq
but the final term in $P$ (out of the plane of the orbit) oscillates as the spin precesses in the plane of the orbit.
Note that the radial component of $P$ is zero.

The spin-curvature force (\ref{Fspin}) simplifies to
\begin{eqnarray}
\label{FspinP}
F^{\rm (sc)}&=&\gamma\zeta_K^2 \left\{\left[2S_{\hat t\hat r}+\nu S_{\hat r\hat \phi}\right]e_{\hat r}-\left[S_{\hat t\hat \theta}+2\nu S_{\hat \theta\hat \phi}\right]e_{\hat \theta}\right\}\nonumber\\
&=&3\gamma^2\nu \zeta_K^2 (S_U^2 e_{\hat r}-S_U^1 e_{\hat \theta})
\ ,
\end{eqnarray}
while the term on the left hand side of Eq.~(\ref{papcoreqs1}) can be written as
\begin{eqnarray}
\label{motradP}
\frac{DP}{\rmd \tau_U}&=&[m\kappa-m_s\tau_1]e_{\hat r}-\gamma\zeta_K^2 \left[S_{\hat t\hat \theta}+2\nu S_{\hat \theta\hat \phi}\right]e_{\hat \theta}\nonumber\\
&=&[m\kappa-m_s\tau_1]e_{\hat r}-3\gamma^2 \nu \zeta_K^2 S_U^1\, e_{\hat \theta}\ .
\end{eqnarray} 
The force balance equation (\ref{baleq}) reduces to 
\begin{eqnarray}
\label{baleqP}
ma(U)_{\hat r}&=&F^{\rm(so)}_{\hat r} + F^{\rm (sc)}_{\hat r}\ , \nonumber\\
ma(U)_{\hat \theta}&=&0=F^{\rm(so)}_{\hat \theta} + F^{\rm (sc)}_{\hat \theta}\ ,
\end{eqnarray}
where
\begin{eqnarray}\fl\quad
&&ma(U)_{\hat r}
= m\kappa\ , \ \, 
F^{\rm(so)}_{\hat r}
= -m_s\left(\frac{DE_{\hat\phi}}{d\tau_{U}}\right)_{\hat r}=m_s\tau_1\ , \ \,
\nonumber\\ 
&&F^{\rm (sc)}_{\hat r}
=3\gamma^2\nu \zeta_K^2 S_U^3
\ ,
\quad
F^{\rm (sc)}_{\hat \theta}
= -F^{\rm(so)}_{\hat \theta}
= -3\gamma^2\nu \zeta_K^2 S_U^1
\ .
\end{eqnarray}

\subsection{The Tulczyjew (T) supplementary conditions}

The T supplementary conditions imply from (\ref{Ps})
\begin{eqnarray}\fl\quad
\label{Tcondsold}
0&=&S_{\hat t\hat \theta}\frac{\rmd S_{\hat t\hat \theta }}{\rmd \tau_U}+S_{\hat t\hat r}\frac{\rmd S_{\hat t\hat r }}{\rmd \tau_U}+\gamma^2S_{\hat t\hat \phi}\frac{\rmd S_{\hat t\hat \phi }}{\rmd \tau_U}+m\gamma^2\nu S_{\hat t\hat \phi}\nonumber\\
\fl\quad
&&-\gamma\frac{\zeta_K}{\nu_K}\{[\nu S_{\hat t\hat r}S_{\hat t\hat \phi}+\nu_K^2S_{\hat t\hat \theta}S_{\hat r\hat \phi}]-\gamma^2S_{\hat t\hat \phi}[\nu S_{\hat t\hat r}-\nu_K^2S_{\hat r\hat \phi}]\}\ , \nonumber\\
\fl\quad
0&=&S_{\hat r\hat \theta}\frac{\rmd S_{\hat t\hat \theta }}{\rmd \tau_U}-\gamma^2[\nu S_{\hat t\hat r}-S_{\hat r\hat \phi}]\frac{\rmd S_{\hat t\hat \phi}}{\rmd \tau_U}+\gamma\frac{\zeta_K}{\nu_K}[S_{\hat t\hat r}^2-\nu_K^2 S_{\hat r\hat \theta}^2]\nonumber\\
\fl\quad
&&-\gamma^3\frac{\zeta_K}{\nu_K}\left[S_{\hat t\hat r}^2-\nu(1+\nu_K^2)S_{\hat t\hat r}S_{\hat r\hat \phi}+\nu_K^2S_{\hat r\hat \phi}^2\right]-m\gamma^2[S_{\hat t\hat r}-\nu S_{\hat r\hat \phi}]\ , \nonumber\\
\fl\quad
0&=&\gamma^2[\nu S_{\hat t\hat \theta}-S_{\hat \theta \hat \phi}]\left[\frac{\rmd S_{\hat t\hat \phi}}{\rmd \tau_U}-\gamma\nu_K\zeta_K S_{\hat r\hat \phi}\right]+S_{\hat r\hat \theta}\frac{\rmd S_{\hat t\hat r }}{\rmd \tau_U}\nonumber\\
\fl\quad
&&+\gamma^2[S_{\hat t\hat \theta}-\nu S_{\hat \theta \hat \phi}]\left[m+\gamma\frac{\zeta_K}{\nu_K}S_{\hat t\hat r}\right]-\gamma\frac{\zeta_K}{\nu_K}[S_{\hat t\hat r}S_{\hat t\hat \theta}+\nu S_{\hat r\hat \theta}S_{\hat t\hat \phi}]\ , \nonumber\\
\fl\quad
0&=&\gamma^2\nu S_{\hat t\hat \phi}\frac{\rmd S_{\hat t\hat \phi }}{\rmd \tau_U}+S_{\hat \theta\hat \phi}\frac{\rmd S_{\hat t\hat \theta }}{\rmd \tau_U}+S_{\hat r\hat \phi}\frac{\rmd S_{\hat t\hat r }}{\rmd \tau_U}+\gamma^3\frac{\zeta_K}{\nu_K}S_{\hat t\hat \phi}[S_{\hat t\hat r}-\nu\nu_K^2S_{\hat r\hat \phi}]\nonumber\\
\fl\quad
&&+m\gamma^2S_{\hat t\hat \phi}-\gamma\frac{\zeta_K}{\nu_K}[\nu_K^2S_{\hat r\hat \theta}S_{\hat \theta\hat \phi}+S_{\hat t\hat \phi}(S_{\hat t\hat r}+\nu S_{\hat r\hat \phi})]\ .
\end{eqnarray}

By solving for the first derivatives, a straightforward calculation shows that the above set of equations simplifies to
\begin{eqnarray}
\label{Tcond1}
0&=&\frac{\rmd S_{\hat t\hat r}}{\rmd \tau_U}+m\frac{S_{\hat t\hat \theta}}{S_{\hat r\hat \theta}}-\gamma\nu\frac{\zeta_K}{\nu_K} S_{\hat t \hat \phi}\ , \\
\label{Tcond2}
0&=&\frac{\rmd S_{\hat t\hat \phi}}{\rmd \tau_U}+m\nu+\gamma\frac{\zeta_K}{\nu_K}[\nu S_{\hat t\hat r}-\nu_K^2 S_{\hat r\hat \phi}]\ , \\
\label{Tcond3}
0&=&\frac{\rmd S_{\hat t\hat \theta}}{\rmd \tau_U}-m\frac{S_{\hat t\hat r}}{S_{\hat r\hat \theta}}-\gamma\nu_K\zeta_K S_{\hat r \hat \theta}\ , \\
\label{Tcond4}
0&=&\frac{m}{\gamma S_{\hat r\hat \theta}}[S_{\hat t\hat \theta}S_{\hat r\hat \phi}-S_{\hat r\hat \theta}S_{\hat t\hat \phi}-S_{\hat t\hat r}S_{\hat \theta \hat \phi}]\ ,
\end{eqnarray}
provided that $S_{\hat r\hat \theta}\not=0$ is assumed.
Substituting Eqs.~(\ref{eqs1b}), (\ref{massvar}) and then Eqs.~(\ref{eqm2c}), (\ref{eqm4d}) into Eq.~(\ref{Tcond2}) (see the equations listed in \ref{appa3}) leads to
\beq
\label{Tcond2b}
0=S_{\hat r\hat \phi}-\frac{\nu_K}{\zeta_K}\left[\gamma\nu c_m-\frac{\nu_K}{\zeta_K}\frac{c_0}{\nu^2-\nu_K^2}\right]\ . 
\eeq
Substituting Eq.~(\ref{eqm4e}) into Eq.~(\ref{Tcond2b}) leads to
\beq
c_1=c_2=c_3=c_4=0\ , \quad c_0=\frac{\gamma\nu}{\gamma_K^2}\frac{\zeta_K}{\nu_K}\left[1+\frac{1}{\gamma^2}\frac{1}{2+\nu_K^2}\right]c_m\ ,
\eeq
implying that $S_{\hat t\hat \phi}=0$, and $S_{\hat t\hat r}$, $S_{\hat r\hat \phi}$ are constant, from Eq.~(\ref{eqs1c}), and Eqs.~(\ref{eqm4e}) and (\ref{eqm2d}) respectively.  
But from Eq.~(\ref{Tcond1}) it follows that $S_{\hat t\hat \theta}=0$, or $A=B=C=D=0$, from Eq.~(\ref{stthsol}), so that $S_{\hat \theta\hat \phi}=0$ 
and $S_{\hat r\hat \theta}=0$ as well, from Eqs.~(\ref{sthphisol}) and  (\ref{srthsol}) respectively. This contradicts the assumption $S_{\hat r\hat \theta}\not=0$ so only the case $S_{\hat r\hat \theta}=0$ remains to be considered. 

If $S_{\hat r\hat \theta}=0$, the set of equations (\ref{Tcondsold}) reduces to
\begin{eqnarray}
\label{Tcond1c2}
0&=&\frac{\rmd S_{\hat t\hat r}}{\rmd \tau_U}-\left[\frac{m}{\nu S_{\hat t\hat r}-S_{\hat r\hat \phi}}+\gamma\nu\frac{\zeta_K}{\nu_K}\right]S_{\hat t\hat \phi}\ , \\
\label{Tcond2c2}
0&=&\frac{\rmd S_{\hat t\hat \phi}}{\rmd \tau_U}-m\frac{\nu S_{\hat r\hat \phi}-S_{\hat t\hat r}}{\nu S_{\hat t\hat r}-S_{\hat r\hat \phi}}+\gamma\frac{\zeta_K}{\nu_K}[\nu S_{\hat t\hat r}-\nu_K^2S_{\hat r\hat \phi}]\ , 
\end{eqnarray}
provided that $\nu S_{\hat t\hat r}-S_{\hat r\hat \phi}\not=0$.
Substituting Eqs.~(\ref{eqs1b}), (\ref{massvar}) and then Eqs.~(\ref{eqm2c}), (\ref{eqm4d}) into Eq.~(\ref{Tcond2c2}) we obtain
\beq\fl\,
\label{Tcond2c2b}
0=\zeta_K^2(\nu^2-\nu_K^2)[\nu_K^2S_{\hat t\hat r}^2-S_{\hat r\hat \phi}^2]+\nu_K^2c_0[\nu S_{\hat t\hat r}-S_{\hat r\hat \phi}]-\gamma\nu_K\zeta_Kc_m(\nu^2-\nu_K^2)[S_{\hat t\hat r}-\nu S_{\hat r\hat \phi}]\,. 
\eeq
Substituting Eq.~(\ref{eqm4e}) into Eq.~(\ref{Tcond2c2b}) then gives
\beq
c_1=c_2=c_3=c_4=0\ , 
\eeq
implying
\beq
\label{vinc1}
S_{\hat t\hat \phi}=0\ , \qquad \frac{\rmd S_{\hat t\hat r}}{\rmd \tau_U}=0=\frac{\rmd S_{\hat r\hat \phi }}{\rmd \tau_U}\ , 
\eeq
from Eq.~(\ref{eqs1c}), and Eqs.~(\ref{eqm4e}) and (\ref{eqm2d}) respectively. 
Hence, Eq.~(\ref{Tcond1c2}) is identically satisfied; moreover
\begin{eqnarray}\fl\,
\label{c0pmsol}
c_0^{(\pm)}&=&\frac{c_m}2\frac{\gamma\nu}{\gamma_K^2}\frac{\zeta_K}{\nu_K}\bigg\{[2-\nu_K^2(1-5\nu_K^2)](\nu^2-\nu_K^2)^2+3\nu_K^2\{\nu^2[3-\nu_K^2(7-\nu_K^2)]\nonumber\\
\fl\,
&&+\nu_K^4(4-\nu_K^2)\}\pm\frac{\nu_K}{\gamma^2\nu}\left[\frac{\nu^2}{\gamma_K^2}+\nu_K^2(2+\nu_K^2)\right][\nu^2(13\nu_K^2+4\nu^2)-8\nu_K^4]^{1/2}\bigg\}\cdot\nonumber\\
\fl\,
&&\cdot\bigg\{\left[\frac{\nu^2}{\gamma_K^2}+\nu_K^2(2+\nu_K^2)\right]^2-9\nu^2\nu_K^4\bigg\}^{-1}\ .
\end{eqnarray}
Next substituting Eqs.~(\ref{eqm4e}) and (\ref{eqm2d}) and then Eq.~(\ref{c0pmsol}) into Eq.~(\ref{massvar}) leads to
\begin{eqnarray}\fl\,
\label{cmpmsol}
c_m^{(\pm)}&=&-\frac{m}2\frac{\{\nu_K^2\nu^4-\nu^2[1-\nu_K^2(3-\nu_K^2)]\}^{-1}}{\nu^2+2\nu_K^2}\bigg\{2\frac{\nu^6}{\gamma_K^2}-\nu^4[1-\nu_K^2(3+\nu_K^2)]\nonumber\\
\fl\,
&&+\nu^2\nu_K^2[2-\nu_K^2(18-\nu_K^2)]+4\nu_K^4(2+\nu_K^2)\nonumber\\
\fl\,
&&\pm\frac{\nu}{\nu_K}\{\nu^2[1-\nu_K^2(3+\nu_K^2)]+\nu_K^2(2+\nu_K^4)\}[\nu^2(13\nu_K^2+4\nu^2)-8\nu_K^4]^{1/2}\bigg\}\ . 
\end{eqnarray}
Finally substituting Eqs.~(\ref{c0pmsol}) and (\ref{cmpmsol}) into Eqs.~(\ref{eqm4e}) and (\ref{eqm2d}), we obtain expressions for the only nonvanishing components of the spin tensor
\begin{eqnarray}\fl\quad
\label{strsolT}
S_{\hat t\hat r}&=&\frac{m}{2\gamma}\frac{\nu_K}{\zeta_K}\frac{\nu^2-\nu_K^2}{\nu^2+2\nu_K^2}\frac{\nu^2(2-3\nu_K^2)+4\nu_K^2\pm\nu\nu_K[\nu^2(13\nu_K^2+4\nu^2)-8\nu_K^4]^{1/2}}{\nu_K^2(\nu^4-2)-\nu^2[1-\nu_K^2(3-\nu_K^2)]}\ , \\
\fl\quad
\label{srphisolT}
S_{\hat r\hat \phi}&=&\frac{m}{2\gamma}\frac{\nu_K}{\zeta_K}\frac{\{\nu_K^2(\nu^4-2)-\nu^2[1-\nu_K^2(3-\nu_K^2)]\}^{-1}}{\nu^2+2\nu_K^2}\{\nu\nu_K[(\nu^2+2\nu_K^2)(4\nu^2-3-\nu_K^2)\nonumber\\
\fl\quad
&&-2(\nu^2-\nu_K^2)^2]\pm[\nu^2(1-3\nu_K^2)-2\nu_K^2][\nu^2(13\nu_K^2+4\nu^2)-8\nu_K^4]^{1/2}\}\ ,
\end{eqnarray}
which are in agreement with the condition $\nu S_{\hat t\hat r}-S_{\hat r\hat \phi}\not=0$ assumed above. This solution, having constant spin components, was already found in previous work \cite{bdfg1}.

Eq.~(\ref{vinc1}) together with the fact that $S_{\hat t\hat \theta}=0$, $S_{\hat r\hat \theta}=0$ show that the total 4-momentum $P$ (see Eq.~(\ref{Ptot})) also lies in the cylinder of the circular orbit 
\beq
\label{PtotT}
P=mU+m_sE_{\hat \phi}\ ,
\eeq
with
\beq
\label{msT}
m_s=\gamma\frac{\zeta_K}{\nu_K}(\nu S_{\hat t\hat r}-\nu_K^2 S_{\hat r\hat \phi})\ .
\eeq
It can therefore be written in the form $P=\mu \, U_p$ with
\begin{equation}\fl\quad 
\label{PtotTcirc}
U_p=\gamma_p\, [e_{\hat t}+\nu_p e_{\hat \phi}]\ , \qquad \nu_p=\frac{\nu+m_s/m}{1+\nu m_s/m}\ ,\qquad \mu=\frac{\gamma}{\gamma_p}(m+\nu m_s)\ ,
\end{equation}
where $\gamma_p=(1-\nu_p^2)^{-1/2}$, provided that $m+\nu m_s\not=0$. 
The T supplementary conditions can then be written as 
\beq
\label{Tcondsnew}
S_{\hat t \hat \phi}=0\ , \qquad S_{\hat r \hat t}+S_{\hat r \hat \phi}\nu_p=0\ , \qquad S_{\hat \theta  \hat t}+S_{\hat \theta  \hat \phi}\nu_p=0\ , 
\eeq
the last condition being identically satisfied, and with the equivalent azimuthal velocity $\nu_p$ given by
\begin{equation}
\label{nupsol}
\nu_p^{(\pm)}=\frac12\frac{\nu_K}{\nu^2+2\nu_K^2}\{3\nu\nu_K\pm[\nu^2(13\nu_K^2+4\nu^2)-8\nu_K^4]^{1/2}\}\ ,
\end{equation}
from Eqs.~(\ref{strsolT}) and (\ref{srphisolT}).
The reality condition of (\ref{nupsol}) requires that $\nu$ takes values outside the interval $(-{\hat \nu},{\hat \nu})$, 
with ${\hat \nu}=\nu_K\sqrt{2}\sqrt{-13+3\sqrt{33}}/4\simeq 0.727 \nu_K$; moreover, 
the timelike condition for $|\nu_p| <1$ is satisfied for all values of $\nu$ outside the same interval.

From (\ref{Tcondsnew}) the spin vector orthogonal to $U_p$ is just
$ {\gamma_p}^{-1} S_{\hat r \hat \phi} E_{\hat \theta}$.
The spin-curvature force (\ref{Fspin}) turns out to be radially directed
\begin{eqnarray}
\label{FspinT}
F^{\rm (sc)}
 = \gamma\zeta_K^2 \left[2S_{\hat t\hat r}+\nu S_{\hat r\hat \phi}\right]e_{\hat r}\ .
\end{eqnarray}
The term on the left hand side of Eq.~(\ref{papcoreqs1}) can be written as 
\beq
\label{motradT}
\frac{DP}{\rmd \tau_U}=[m\kappa-m_s\tau_1]e_{\hat r}\ ,
\eeq 
so that the balance equation (\ref{baleq}) reduces to 
\beq
\label{baleqT}
ma(U)_{\hat r}=F^{\rm(so)}_{\hat r} + F^{\rm (sc)}_{\hat r}\ , 
\eeq
where
\beq\fl\quad
ma(U)_{\hat r}=m\kappa\ , \quad F^{\rm(so)}_{\hat r}=m_s\tau_1\ , \quad 
F^{\rm (sc)}_{\hat r}=\gamma\zeta_K^2\left[2S_{\hat t\hat r}+\nu S_{\hat r\hat \phi}\right]\ .
\eeq

\section{Conclusions}

Spinning test particles in  circular motion around a Schwarzschild black hole have been discussed in the framework of the 
Mathisson-Papapetrou approach supplemented by the usual standard conditions.
One finds that apart from very special (and indeed artificially constrained) orbits where the spin tensor is closely matched to the curvature and torsion properties of the world line of the test particle or the static observer spin vector is constant and orthogonal to the plane of the orbit, the assumption of circular motion is not compatible with these equations. Indeed even in the former case, the test particle assumption is then violated except in the limit of massless particles following null geodesics, where the spin vector must be aligned with the direction of motion from general considerations.
The spin-curvature force generically forces the motion away from circular orbits, so one needs a much more complicated machinery to attempt to study explicit solutions of this problem, solutions which must break the stationary axisymmetry.

\appendix
\section{Derivation of the solutions of the equations of motion}

This Appendix derives the solutions of the equations of motion alone  Eqs.~(\ref{eqs1})--(\ref{eqs3}) and (\ref{eqm1})--(\ref{eqm4}) for the spin tensor along a circular orbit without supplementary conditions imposed. Standard elimination and differentiation techniques are used to find decoupled second order linear constant coefficient equations for certain spin components, from which one may calculate the remaining spin components that do not already satisfy decoupled first order such equations.

\subsection{The $\nu=0$ case}

When the particle is at rest $\nu=0$ relative to the static observers, these equations reduce to
\begin{eqnarray}
\label{eq1nueq0} 
0&=&\frac{\rmd S_{\hat r\hat \phi}}{\rmd \tau_U}-\nu_K\zeta_K S_{\hat t\hat \phi}\ , \\
\label{eq2nueq0}
0&=&\frac{\rmd S_{\hat \theta\hat \phi}}{\rmd \tau_U}\ , \\
\label{eq3nueq0}
0&=&\frac{\rmd S_{\hat r\hat \theta}}{\rmd \tau_U}-\nu_K\zeta_K S_{\hat t\hat \theta}\ , \\
\label{eq4nueq0}
0&=&\frac{\rmd m}{\rmd \tau_U}+\nu_K\zeta_K \frac{\rmd S_{\hat t\hat r}}{\rmd \tau_U}\ , \\
\label{eq5nueq0}
0&=&\frac{\rmd^2 S_{\hat t\hat r}}{\rmd \tau_U^2}-2\zeta_K^2 S_{\hat t\hat r}+m\nu_K\zeta_K\ , \\
\label{eq6nueq0}
0&=&\frac{\rmd^2S_{\hat t\hat \theta}}{\rmd \tau_U^2}+\zeta_K^2 S_{\hat t\hat \theta}-\zeta_K\nu_K\frac{\rmd S_{\hat r\hat \theta}}{\rmd \tau_U}\ , \\
\label{eq7nueq0}
0&=&\frac{\rmd^2S_{\hat t\hat \phi}}{\rmd \tau_U^2}+\zeta_K^2 S_{\hat t\hat \phi}-\zeta_K\nu_K\frac{\rmd S_{\hat r\hat \phi}}{\rmd \tau_U}\ . 
\end{eqnarray}

Eq.~(\ref{eq2nueq0}) implies that $S_{\hat \theta\hat \phi}=c_1$ is constant.
Solving Eq.~(\ref{eq3nueq0}) for $\rmd S_{\hat r\hat \theta}/\rmd \tau_U$, and substituting the result into Eq.~(\ref{eq6nueq0}) leads to the decoupled equation
\beq
\label{eq6bnueq0}
0=\frac{\rmd^2S_{\hat t\hat \theta}}{\rmd \tau_U^2}+\omega_0^2 S_{\hat t\hat \theta}\ , \qquad \omega_0=\frac{\zeta_K}{\gamma_K}=\sqrt{\frac{M(r-3M)}{r^3(r-2M)}}\ .
\eeq
Similarly solving Eq.~(\ref{eq1nueq0}) for $\rmd S_{\hat r\hat \phi}/\rmd \tau_U$, and substituting the result into Eq.~(\ref{eq7nueq0}) leads to an analogous decoupled equation
\beq
\label{eq7bnueq0}
0=\frac{\rmd^2S_{\hat t\hat \phi}}{\rmd \tau_U^2}+\omega_0^2 S_{\hat t\hat \phi}\ .
\eeq
Eq.~(\ref{eq4nueq0}) leads immediately to $m=-\nu_K\zeta_K S_{\hat t\hat r}+c_m$
and substituting this into  (\ref{eq5nueq0}) yields
\beq\fl\quad
\label{eq5bnueq0}
0=\frac{\rmd^2S_{\hat t\hat r}}{\rmd \tau_U^2}+\omega_1^2 S_{\hat t\hat r}+\nu_K\zeta_K c_m\ , \qquad \omega_1=\zeta_K(2+\nu_K^2)^{1/2}=\sqrt{\frac{M(2r-3M)}{r^3(r-2M)}}\ .
\eeq
The three second order constant coefficient Eqs.~(\ref{eq6bnueq0}), (\ref{eq7bnueq0}) and (\ref{eq5bnueq0})  are easily integrated, from which expressions for the remaining components of the spin tensor are then obtained from Eqs.~(\ref{eq1nueq0}) and  (\ref{eq3nueq0}). These have either oscillatory or exponential solutions depending on whether the squared frequencies in Eqs.~(\ref{eq6bnueq0}) and (\ref{eq5bnueq0}) are positive or negative, or linear solutions when zero. 
This distinguishes the two intervals $2M<r<3M$ and $r>3M$, whose corresponding solutions are given by Eqs.~(\ref{solnueq0I}) and (\ref{solnueq0II}) respectively.
The special case $r=3M$ can be easily discussed by setting $\nu_K=1$ and $\zeta_K=\sqrt{3}/(9M)$ in Eqs.~(\ref{eq1nueq0})--(\ref{eq7nueq0}). The corresponding solution is given by Eq.~(\ref{solnueq0req3M}).

\subsection{The $\nu=\pm\nu_K$ case}

When the particle moves along a geodesic with $\nu=\pm\nu_K$, Eqs.~(\ref{eqs1})--(\ref{eqs3}) and (\ref{eqm1})--(\ref{eqm4}) simplify to
\begin{eqnarray}
\label{eq1nuK} 
0&=&\frac{\rmd S_{\hat r\hat \phi}}{\rmd \tau_U}\mp\nu_K \frac{\rmd S_{\hat t\hat r}}{\rmd \tau_U}\ , \\
\label{eq2nuK}
0&=&\frac{\rmd S_{\hat \theta\hat \phi}}{\rmd \tau_U}\mp\nu_K \frac{\rmd S_{\hat t\hat \theta}}{\rmd \tau_U}
       \mp\frac{\zeta_K}{\gamma_K}S_{\hat r\hat \theta}\ , \\
\label{eq3nuK}
0&=&\frac{\rmd S_{\hat r\hat \theta}}{\rmd \tau_U}\pm\zeta_K\gamma_K [S_{\hat \theta\hat \phi}
  \mp\nu_K S_{\hat t\hat \theta}]\ , \\
\label{eq4nuK}
0&=&\frac{\rmd^2S_{\hat t\hat \phi}}{\rmd \tau_U^2}\pm\frac1{\nu_K}\frac{\rmd m}{\rmd \tau_U}
  \pm2\zeta_K\gamma_K \left[\frac{\rmd S_{\hat t\hat r}}{\rmd \tau_U}\mp\nu_K\frac{\rmd S_{\hat r\hat \phi}}{\rmd \tau_U}\right]\ , \\
\label{eq5nuK}
0&=&\frac{\rmd^2 S_{\hat t\hat r}}{\rmd \tau_U^2}\mp\nu_K\frac{\rmd^2 S_{\hat r\hat \phi}}{\rmd \tau_U^2}
  \mp2\frac{\zeta_K}{\gamma_K}\frac{\rmd S_{\hat t\hat \phi}}{\rmd \tau_U}-3\zeta_K^2 S_{\hat t\hat r}\ , \\
\label{eq6nuK}
0&=&\frac{\rmd^2S_{\hat t\hat \theta}}{\rmd \tau_U^2}
  \mp\nu_K\frac{\rmd^2S_{\hat \theta\hat \phi}}{\rmd \tau_U^2}+\zeta_K^2 [S_{\hat t\hat \theta}
  \pm 2\nu_K S_{\hat \theta\hat \phi}]\ , \\
\label{eq7nuK}
0&=&\frac{\rmd^2S_{\hat t\hat \phi}}{\rmd \tau_U^2}\pm\nu_K\frac{\rmd m}{\rmd \tau_U}
  \pm2\zeta_K\gamma_K \left[\frac{\rmd S_{\hat t\hat r}}{\rmd \tau_U}
  \mp\nu_K\frac{\rmd S_{\hat r\hat \phi}}{\rmd \tau_U}\right]\ .
\end{eqnarray}

The difference of Eqs.~(\ref{eq4nuK}) and (\ref{eq7nuK}) leads to $\rmd m/\rmd \tau_U=0$, so
$m=c_m$ is constant, which then implies from the same equations that
\beq
\label{eq7bnuK}
0=\frac{\rmd^2S_{\hat t\hat \phi}}{\rmd \tau_U^2}\pm2\zeta_K\gamma_K \left[\frac{\rmd S_{\hat t\hat r}}{\rmd \tau_U}\mp\nu_K\frac{\rmd S_{\hat r\hat \phi}}{\rmd \tau_U}\right]\ .
\eeq
Integration of Eq.~(\ref{eq1nuK}) yields
\beq
\label{eq1bnuK}
S_{\hat r\hat \phi}=\pm\nu_K S_{\hat t\hat r}+c_1\ ,
\eeq
and using this to replace $S_{\hat r\hat \phi}$ in Eq.~(\ref{eq7bnuK}) and then integrating gives
\beq
\label{eq7cnuK}
\frac{\rmd S_{\hat t\hat \phi}}{\rmd \tau_U} \pm2\frac{\zeta_K}{\gamma_K} S_{\hat t\hat r}=c_2\ .
\eeq
Using  Eqs.~(\ref{eq1bnuK}) and  (\ref{eq7cnuK}) to replace $S_{\hat r\hat \phi}$ 
and $\rmd S_{\hat t\hat \phi}/\rmd \tau_U$ in
Eq.~(\ref{eq5nuK}) then leads to the decoupled second order equation
\beq\fl\quad
\label{eq5bnuK}
0=\frac{\rmd^2 S_{\hat t\hat r}}{\rmd \tau_U^2}+\omega_2^2 S_{\hat t\hat r}\pm2\gamma_K\zeta_Kc_2
\ , \qquad 
\omega_2=\zeta_K(4-3\gamma_K^2)^{1/2}=\sqrt{\frac{M(r-6M)}{r^3(r-3M)}}\ .
\eeq
Taking the $\tau_U$ derivative of Eq.~(\ref{eq2nuK}) and using it and Eq.~(\ref{eq3nuK}) in Eq.~(\ref{eq6nuK})
leads to the second order equation
\beq\fl\quad
\label{eq6bnuK}
0=\frac{\rmd^2 S_{\hat t\hat \theta}}{\rmd \tau_U^2}
  +\omega_3^2 S_{\hat t\hat \theta}+3\zeta_K^2\gamma_K^2\nu_K S_{\hat \theta\hat \phi}
\ , \qquad 
\omega_3=\zeta_K=\left(\frac{M}{r^3}\right)^{1/2}\ .
\eeq
Finally using Eq.~(\ref{eq6bnuK}) and Eq.~(\ref{eq3nuK}) in the equation obtained by taking the $\tau_U$ derivative of Eq.~(\ref{eq2nuK}) gives a second decoupled second order equation 
\beq\fl\quad
\label{eq2bnuK}
0=\frac{\rmd^2 S_{\hat \theta\hat \phi}}{\rmd \tau_U^2}+\omega_4^2 S_{\hat \theta\hat \phi}
\ , \qquad 
\omega_4=\zeta_K(3\gamma_K^2-2)^{1/2}=\sqrt{\frac{M}{r^3(r-3M)}}\ .
\eeq
Integrating the two decoupled second order equations (\ref{eq5bnuK}) and (\ref{eq2bnuK}), one can then integrate Eq.~(\ref{eq6bnuK}) too. The remaining components of the spin tensor are then determined by 
Eqs.~(\ref{eq3nuK}), (\ref{eq1bnuK}), and  (\ref{eq7cnuK}).
Note that the frequencies $\omega_3$ and $\omega_4$ agree only for $\gamma_K=1$, or $\nu_K=0$, which would imply $M=0$: so this special case is not relevant.

Now the two intervals $3M<r<6M$ and $r>6M$ have differing signs for the squared angular velocities, and the corresponding solutions are given by Eqs.~(\ref{solnuKII}) and (\ref{solnuKIII}) respectively.
The special case $r=6M$ can be easily handled as well, setting $\nu_K=1/2$ and $\zeta_K=\sqrt{6}/(36M)$ in Eqs.~(\ref{eq1nuK})--(\ref{eq7nuK}). The corresponding solution is given by Eq.~(\ref{solnuKreq6M}).

\subsection{The general case: $\nu\not=0$ and $\nu\not=\pm\nu_K$}
\label{appa3}

Here we deal with the general form of Eqs.~(\ref{eqs1})--(\ref{eqs3}) and (\ref{eqm1})--(\ref{eqm4}).
Solving Eqs.~(\ref{eqs1})--(\ref{eqs3}) for their first terms and substituting these derivative terms into Eqs.~(\ref{eqm1})--(\ref{eqm4}), one finds
\begin{eqnarray}\fl\,
\label{eqm1bis} 
\frac{\rmd^2S_{\hat t\hat \phi}}{\rmd \tau_U^2}
&=&\frac{\gamma}{\nu}\frac{\zeta_K}{\nu_K}[(2\nu_K^2-1)\nu^2-\nu_K^2]\frac{\rmd S_{\hat t\hat r}}{\rmd \tau_U}
-\gamma^2\zeta_K^2(\nu^2-\nu_K^2)S_{\hat t\hat \phi}-\frac1{\nu}\frac{\rmd m}{\rmd \tau_U}
\ , \\
\fl\, 
\label{eqm2bis}
\frac{\rmd^2S_{\hat t\hat r}}{\rmd \tau_U^2}
&=&-\gamma^3\nu\frac{\zeta_K}{\nu_K}[\nu^2+\nu_K^2-2]\frac{\rmd S_{\hat t\hat \phi}}{\rmd \tau_U}
-\gamma^4\frac{\zeta_K^2}{\nu_K^2}[(3\nu_K^2-1)\nu^2-2\nu_K^2]S_{\hat t\hat r} 
\nonumber\\
\fl\,
&&-\gamma^4\nu\zeta_K^2(\nu^2-\nu_K^2)S_{\hat r\hat \phi} +m\gamma^3\frac{\zeta_K}{\nu_K}(\nu^2-\nu_K^2)
\ , \\
\fl\,
\label{eqm3bis}
\frac{\rmd^2S_{\hat t\hat \theta}}{\rmd \tau_U^2}
&=&\frac{\gamma^3}{\gamma_K^2}\frac{\zeta_K}{\nu_K}\nu^2\frac{\rmd S_{\hat r\hat \theta}}{\rmd \tau_U}
-\frac{\gamma^4}{\gamma_K^2}\zeta_K^2S_{\hat t\hat \theta}
+\gamma^4\nu\frac{\zeta_K^2}{\nu_K^2}[(1+2\nu_K^2)\nu^2-3\nu_K^2]S_{\hat \theta\hat \phi}
\ , \\
\fl\, 
\label{eqm4bis}
\frac{\rmd^2S_{\hat t\hat \phi}}{\rmd \tau_U^2}
&=&\gamma\nu\frac{\zeta_K}{\nu_K}[\nu^2+\nu_K^2-2]\frac{\rmd S_{\hat t\hat r}}{\rmd \tau_U}
-\gamma^2\frac{\zeta_K^2}{\nu_K^2}(\nu^2-\nu_K^2)[\nu^2+\nu_K^2-1]S_{\hat t\hat \phi}-\nu\frac{\rmd m}{\rmd \tau_U}\ . 
\end{eqnarray}

Solving Eqs.~(\ref{eqm1bis}) and (\ref{eqm4bis}) for $\rmd m/\rmd \tau_U$ and $\rmd^2 S_{\hat r\hat \phi}/\rmd \tau_U^2$, one finds
\begin{eqnarray}
\label{eqm1b}
\frac{\rmd m}{\rmd \tau_U}&=&\gamma\frac{\zeta_K}{\nu_K}(\nu^2-\nu_K^2)\left[\frac{\rmd S_{\hat t\hat r}}{\rmd \tau_U}-\gamma\nu\frac{\zeta_K}{\nu_K}S_{\hat t\hat \phi}\right]\ ,  \\
\label{eqm4b}
\frac{\rmd^2S_{\hat t\hat \phi}}{\rmd \tau_U^2}&=&-2\frac{\gamma\nu}{\gamma_K^2}\frac{\zeta_K}{\nu_K}\frac{\rmd S_{\hat t\hat r}}{\rmd \tau_U}+\frac{\gamma^2}{\gamma_K^2}\frac{\zeta_K^2}{\nu_K^2}(\nu^2-\nu_K^2)S_{\hat t\hat \phi}\ .
\end{eqnarray}
Solving Eq.~(\ref{eqs1}) for $S_{\hat t\hat \phi}$, leads to
\beq
\label{eqs1b}
S_{\hat t\hat \phi}
=\frac1{\gamma}\frac{\nu_K}{\zeta_K}\frac1{\nu^2-\nu_K^2}
 \left[\nu\frac{\rmd S_{\hat t\hat r}}{\rmd \tau_U}-\frac{\rmd S_{\hat r\hat \phi}}{\rmd \tau_U}\right]\ .
\eeq
Using Eq.~(\ref{eqs1b}) in Eq.~(\ref{eqm1b}) leads to
\beq
\label{eqm1c}
m=\gamma\frac{\zeta_K}{\nu_K}[\nu S_{\hat r\hat \phi}- \nu_K^2 S_{\hat t\hat r}]+c_m\ .
\eeq
Then using Eqs.~(\ref{eqs1b}) and  (\ref{eqm1c}) in Eq.~(\ref{eqm2bis}) leads to 
\begin{eqnarray}\fl\quad
\label{eqm2b}
0&=&[(1-2\nu_K^2)\nu^2+\nu_K^2]\frac{\rmd^2S_{\hat t\hat r}}{\rmd \tau_U^2}+\nu[\nu^2+\nu_K^2-2]\frac{\rmd^2S_{\hat r\hat \phi}}{\rmd \tau_U^2}+\frac{\gamma^2}{\gamma_K^2}\nu\frac{\zeta_K^2}{\nu_K^2}(\nu^2-\nu_K^2)^2S_{\hat r\hat \phi}\nonumber\\
\fl\quad
&&+\gamma^2\frac{\zeta_K^2}{\nu_K^2}(\nu^2-\nu_K^2)[(1-4\nu_K^2)\nu^2+\nu_K^2(2+\nu_K^2)]S_{\hat t\hat r}
+\gamma\frac{\zeta_K}{\nu_K}(\nu^2-\nu_K^2)^2c_m\ . 
\end{eqnarray}
Using Eq.~(\ref{eqs1b}) in Eq.~(\ref{eqm4b}) leads to 
\beq
\label{eqm4c}
\frac{\rmd^2S_{\hat r\hat \phi}}{\rmd \tau_U^2}-\nu\frac{\rmd^2S_{\hat t\hat r}}{\rmd \tau_U^2}-\frac{\gamma^2}{\gamma_K^2}\frac{\zeta_K^2}{\nu_K^2}(\nu^2-\nu_K^2)[S_{\hat r\hat \phi}+\nu S_{\hat t\hat r}]=c_0\ .
\eeq
Then solving these last two equations for $\rmd^2S_{\hat r\hat \phi}/\rmd \tau_U^2$ and $\rmd^2S_{\hat t\hat r}/\rmd \tau_U^2$ gives
\begin{eqnarray}\fl\quad
\label{eqm2c}
\frac{\rmd^2S_{\hat t\hat r}}{\rmd \tau_U^2}
&=&-\gamma^2\frac{\zeta_K^2}{\nu_K^2}\left[\frac{\nu^2}{\gamma_K^2}-\nu_K^2(2+\nu_K^2)\right] S_{\hat t\hat r}
-2\frac{\gamma^2}{\gamma_K^2}\nu\frac{\zeta_K^2}{\nu_K^2}S_{\hat r\hat \phi}
\nonumber\\
\fl\quad
&&+\gamma^2\nu\frac{\nu^2+\nu_K^2-2}{\nu^2-\nu_K^2}c_0+\gamma^3\frac{\zeta_K}{\nu_K}(\nu^2-\nu_K^2)c_m
\ , \\
\fl\quad
\label{eqm4d}
\frac{\rmd^2S_{\hat r\hat \phi}}{\rmd \tau_U^2}
&=&-\frac{\gamma^2}{\gamma_K^2}\frac{\zeta_K^2}{\nu_K^2}(\nu^2+\nu_K^2)S_{\hat r\hat \phi}
+\gamma^2\nu\zeta_K^2(1+2\nu_K^2)S_{\hat t\hat r}
\nonumber\\
\fl\quad
&&-\gamma^2\frac{\nu^2(1-2\nu_K^2)+\nu_K^2}{\nu^2-\nu_K^2}c_0+\gamma^3\nu\frac{\zeta_K}{\nu_K}(\nu^2-\nu_K^2)c_m\ ,
\end{eqnarray}
whose solution is given by Eqs.~(\ref{eqm4e}) and (\ref{eqm2d}). 
Then substituting these solutions into Eq.~(\ref{eqs1b}), one obtains (\ref{eqs1c}). 

Substituting Eqs.~(\ref{eqs2}) and (\ref{eqs3}) into Eq.~(\ref{eqm3}) gives
\beq 
\label{eqm3b} 
0=\frac{\rmd^2S_{\hat t\hat \theta}}{\rmd \tau_U^2}+\Omega_1^2 S_{\hat t\hat \theta}
+3\gamma^2\nu\zeta_K^2 S_{\hat \theta\hat \phi}
\ , \qquad 
\Omega_1^2=\gamma^2\frac{\zeta_K^2}{\gamma_K^2}\ .
\eeq
Taking the $\tau_U$ derivative of Eq.~(\ref{eqs2}) and using Eqs.~(\ref{eqm3b}) and (\ref{eqs3}), one gets
\beq
\label{eqs2b}  
0=\frac{\rmd^2S_{\hat \theta\hat \phi}}{\rmd \tau_U^2}+\Omega^2 S_{\hat \theta\hat \phi}
\ , \qquad 
\Omega^2=\gamma^2\nu^2\frac{\zeta_K^2}{\nu_K^2}(1+2\nu_K^2)\ .
\eeq
This is easily integrated to give the solution (\ref{sthphisol}).
The remaining components of the spin can be obtained directly from Eqs.~(\ref{eqm3b}) and (\ref{eqs3})
yielding  Eqs.~(\ref{stthsol}) and (\ref{srthsol}). 
The frequencies $\Omega$ and $\Omega_1$ agree for the particular value of the azimuthal velocity given by
\beq
\nu_0=\pm\frac{\nu_K}{\gamma_K}(1+2\nu_K^2)^{-1/2}\ .
\eeq
The solution corresponding to this special case is given by Eqs.~(\ref{sthphisolomeqom1})--(\ref{srthsolomeqom1}) together with Eqs.~(\ref{eqm4e})--(\ref{massvar}) evaluated at $\nu=\nu_0$.


\section*{References}

\begin{thebibliography}{00}


\bibitem{math37} 
 Mathisson M 1937
  {\it Acta Phys.\ Polonica} {\bf 6} 167 

 \bibitem{papa51} 
 Papapetrou A 1951
  {\it Proc.\ Roy.\ Soc.\ London} {\bf 209} 248

\bibitem{cori51}
 Corinaldesi E and Papapetrou A 1951
  {\it Proc.\ Roy.\ Soc.\ London} {\bf 209} 259 
 
 \bibitem{pir56} 
 Pirani F 1956
  {\it Acta Phys.\ Polon.} {\bf 15} 389 

\bibitem{tulc59} 
 Tulczyjew W 1959
  {\it Acta Phys.\ Polon.} {\bf 18} 393 

 \bibitem{bdfg1}
  Bini D, de Felice F and Geralico A 2004
   {\it Class.\ Quantum Grav.} {\bf 21} 5427 

\bibitem{idcf1}
 Bini D, Carini P and Jantzen R T 1997
  {\it Int.\ J.\ Mod.\ Phys. D\/} {\bf 6} 1 

 \bibitem{idcf2}
  Bini D, Carini P and Jantzen R T 1997
   {\it Int.\ J.\ Mod.\ Phys. D\/} {\bf 6} 143 

\bibitem{iyer-vish} 
 Iyer B R  and Vishveshwara C V 1993
  {\it Phys.\ Rev.\ D\/} {\bf 48} 5721

 \bibitem{mol}
  M\o ller C 1949
  {\it Commun. Dublin Inst. Adv. Studies} A  {\bf 5} 3 

\bibitem{mashnull} 
 Mashhoon B 1975
  {\it Ann. Phys.} {\bf 89} 254 

 \bibitem{maeda} 
  Suzuki S and Maeda K 1997
   {\it Phys.\ Rev.\ D\/} {\bf 55} 4848 

\bibitem{semerak}
 Semer\'ak O 1999
  {\it MNRAS} {\bf 308} 863 

 \bibitem{hartl1} 
  Hartl M D 2003
   {\it Phys.\ Rev.\ D\/} {\bf 67} 024005 

\bibitem{hartl2} 
 Hartl M D 2003
  {\it Phys.\ Rev.\ D\/} {\bf 67} 104023 


\end{thebibliography}
\end{document}